\documentclass[%
 aip,
 amsmath,amssymb,
 reprint,%
]{revtex4-1}

\usepackage{graphicx}
\usepackage{dcolumn}
\usepackage{bm}

\usepackage[utf8]{inputenc}
\usepackage[T1]{fontenc}
\usepackage{mathptmx}
\usepackage{etoolbox}

\makeatletter
\def\@email#1#2{%
 \endgroup
 \patchcmd{\titleblock@produce}
  {\frontmatter@RRAPformat}
  {\frontmatter@RRAPformat{\produce@RRAP{*#1\href{mailto:#2}{#2}}}\frontmatter@RRAPformat}
  {}{}
}%
\makeatother
\begin{document}

\preprint{AIP/123-QED}

\title[]{A sub-volt near-IR lithium tantalate electro-optic modulator}
\author{Keith Powell}
\altaffiliation[]{keith@luminacorp.com.au}
\affiliation{John A. Paulson School of Engineering and Applied Science, Harvard University, 29 Oxford St., Cambridge, MA 02138 USA}
\author{Dylan Renaud}%
\affiliation{John A. Paulson School of Engineering and Applied Science, Harvard University, 29 Oxford St., Cambridge, MA 02138 USA}
\author{Xudong Li}
\affiliation{John A. Paulson School of Engineering and Applied Science, Harvard University, 29 Oxford St., Cambridge, MA 02138 USA}
\author{Daniel Assumpcao}
\affiliation{John A. Paulson School of Engineering and Applied Science, Harvard University, 29 Oxford St., Cambridge, MA 02138 USA}
\author{C. J. Xin}
\affiliation{John A. Paulson School of Engineering and Applied Science, Harvard University, 29 Oxford St., Cambridge, MA 02138 USA}
\author{Neil Sinclair}
\altaffiliation[]{neils@seas.harvard.edu}
\affiliation{John A. Paulson School of Engineering and Applied Science, Harvard University, 29 Oxford St., Cambridge, MA 02138 USA}
\author{Marko~Lončar}
\altaffiliation[]{loncar@seas.harvard.edu}
\affiliation{John A. Paulson School of Engineering and Applied Science, Harvard University, 29 Oxford St., Cambridge, MA 02138 USA}

\date{\today}

\begin{abstract}
We demonstrate a low-loss integrated electro-optic Mach-Zehnder modulator in thin-film lithium tantalate at 737 nm, featuring a low half-wave voltage-length product of 0.65 V$\cdot$cm, an extinction ratio of 30 dB, low optical loss of 5.3 dB, and a detector-limited bandwidth of 20 GHz. 
A small $<2$ dB DC bias drift relative to quadrature bias is measured over 16 minutes using 4.3 dBm of on-chip power in ambient conditions, which outperforms the 8 dB measured using a counterpart thin-film lithium niobate modulator. 
Finally, an optical loss coefficient of 0.5 dB/cm for a thin-film lithium tantalate waveguide is estimated at 638 nm using a fabricated ring resonator.
\end{abstract}

\maketitle

\section{Introduction}
Thin-film lithium niobate (TFLN) electro-optic (EO) integrated circuits have shown promise in furthering optical science and technology \cite{hu2025integrated,boes2023lithium,zhu2021integrated,xu2020high,desiatov2019ultra}.
Their advantage derives from the combination of large Pockels coefficient ($\sim$30 pm/V) across a wide  wavelength range (LN bandgap is 3.63 eV) and the ability to realize low-loss optical waveguides using conventional nanofabrication techniques (lithography, etching, etc). 
Recently, thin-film lithium tantalate (TFLT) has become commercially available, and has been explored for electro-optic circuits at telecommunication wavelength \cite{powell2024stable,wang2023,DC_EO_LT,shen23,yu_24,nishi_22,wang2024ultrabroadband,zhang2025ultrabroadband}.
This has been primarily motivated by the similar or even improved properties LT has compared to LN: EO coefficient of $r_{33}$\textasciitilde30~pm/V \cite{LT_pockels}, bandgap of 3.93 eV \cite{LT_bandgap}, $23\times$ lower birefringence than LN at 633~nm \cite{LT_good}, $5\times$ lower photorefraction than LN at visible wavelengths \cite{LT_phr,LN_phr}, 2500$\times$ higher optical damage threshold than LN for green light \cite{LT_damage,LN_damage}, and $10\times$ lower RF loss tangent than LN \cite{LT_good,LN_losstan}.

Lower birefringence and reduced effects of charge transport are the main drivers of TFLT photonics. 
For example, it is well-known that photo-induced charge transport along with photovoltaic effects can enhance EO relaxation, which manifests as an uncontrolled variation of the optical phase of light in the crystal due to charge migration.
In particular, the relaxation rate will increase with more applied optical power and can be exacerbated with applied DC or RF field.
This effect reduces the DC stability of electro-optic circuits, such as Mach-Zehnder modulators (MZMs), and has been one of the main challenges faced by TFLN photonics \cite{holzgrafe_TFLN_relaxation}. 
Several methods have been used to mitigate EO relaxation effects and, hence, overcome bias point drifts of EO MZMs. 
These include operation at reduced optical powers (<10 dBm) and use of heaters with the thermo-optic effect to operate at the desired bias point \cite{xu2020high,holzgrafe_TFLN_relaxation}, both of which restrict usability and practicality, or may be infeasible for the application at hand.
Recently, telecommunication-wavelength TFLT MZMs were demonstrated by our group \cite{powell2024stable} and others \cite{yu_24,wang2024ultrabroadband} to have superior performance to TFLN modulators. 
We inferred a slow EO relaxation for on-chip optical powers up to 12.1 dBm \cite{powell2024stable}.

The performance of electro-optic devices in TFLT at other wavelengths has yet to be demonstrated. 
Of particular interest are the visible and near-IR wavelengths \cite{tran2022extending}, which are relevant for applications in imaging \cite{wang2024vivo}, clocks \cite{newman2019architecture}, data centers \cite{cheng2018recent}, displays \cite{morin2021cmos}, spectroscopy \cite{suh2019searching}, and quantum information \cite{bradac2019quantum}, for instance.  
Motivated by these, TFLN MZMs have been developed at 738 nm \cite{renaud2023sub}, from 400-700 nm \cite{Xue23}, at 768 nm by upconversion \cite{Sabatti24}, and recently at 456 nm \cite{AmirSN_456nm}. 
Stronger photo-induced EO relaxation is expected at shorter wavelengths approaching the LN bandgap.

Here we design and fabricate a TFLT MZM operating at a near-IR wavelength of 737 nm that exhibits superior DC bias stability in ambient conditions compared to an equivalent TFLN MZM fabricated with a similar process. 
Specifically, we measure $<2$ dB laser power fluctuations over 16 minutes using TFLT compared to 8 dB over the same timescale using TFLN for an on-chip power of 4.3 dBm when DC biasing the modulator at quadrature.
Furthermore, our TFLT MZMs feature low half-wave voltage length product of 0.65 V$\cdot$cm, a high extinction ratio of 30 dB, low optical loss of 5.3 dB, and a detector-limited bandwidth of 20 GHz. 
Note that interest in 737 nm wavelength is motivated by  silicon vacancy color centers in diamond ( SiV$^-$), one of leading solid state quantum memories \cite{assumpcao2024thin}.
To further assess the optical loss of our TFLT waveguides, we fabricate ring resonators and measure optical quality factors up to $2.8 \times 10^5$  which corresponds to a 0.5 dB/cm propagation coefficient.
This measurement is performed at 638 nm wavelength due to the limited tuning range of our 737 nm-wavelength laser.

\section{Device fabrication}

\begin{figure*}[ht!]
\centering\includegraphics[width=\linewidth]{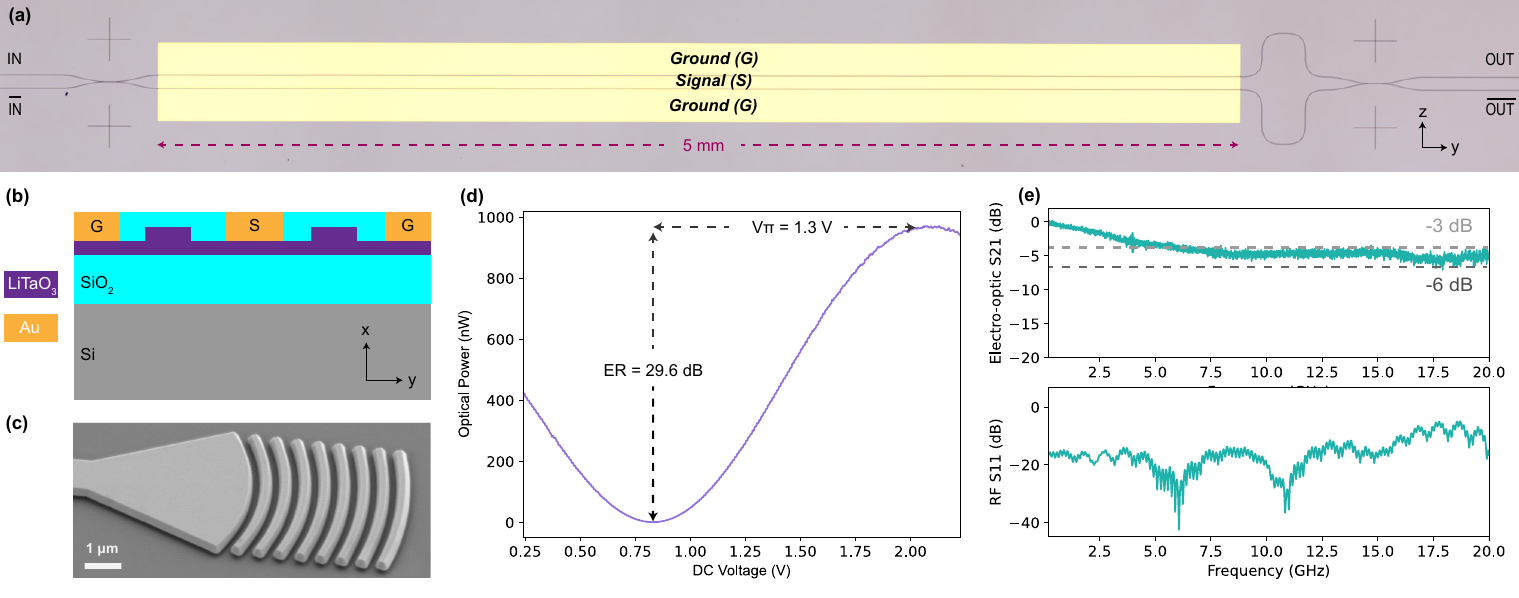}
\caption{Near-IR thin-film lithium tantalate Mach-Zehnder electro-optic modulator (a) Optical micrograph of the fabricated modulator. Crystal axes and electrode details are labeled. (b) Cross-section of the material stack of the modulator with crystal axes indicated. (c) Scanning electron micro-graph showing a grating coupler used to couple light to and from our modulator. (d) Measured transfer function of the modulator with slowly varying applied voltage yields an extinction ratio (ER) of 29.6 dB and half-wave voltage (V$_\pi$) of 1.3 V. (c) Measured electro-optic frequency response (S$_{21}$) is limited by the detector used. Reflected RF power (S$_{11}$) of the modulator transmission line indicates efficient power delivery to the electrodes.}
\label{fig:figure1}
\end{figure*}

An optical microscope image of a fabricated unbalanced MZM is shown in Fig. \ref{fig:figure1}a. 
It consists of a directional coupler as an input beamsplitter and a $L=5$~mm long electrode in the ground-signal-ground configuration followed by another directional coupler at the output.
Grating couplers (not shown in Fig. \ref{fig:figure1}a) are used to couple light on and off the chip to near-IR single-mode fibers.
The directional coupler is chosen to minimize optical insertion loss and is carefully optimized to reach 50:50 splitting. 

A cross-section of the TFLT device stack is shown in Fig. \ref{fig:figure1}b. 
The optical layer of the device is defined using 150 keV electron-beam lithography with 500 nm-thick ma-N2405 resist on top of 200 nm-thick x-cut TFLT-on-SiO$_2$.
The waveguide width is designed to be 600 nm.
The SiO$_2$ layer is 2 $\mu$m-thick and is on a Si substrate.
The TFLT is etched by 100 nm using an Ar$^{+}$-based inductively-coupled plasma reactive-ion etching.
Etch-induced re-deposition is removed using a high-pH solution.
The devices are then annealed in an O$_2$ atmosphere at 520°C for 2 h to mitigate etch-induced imperfections.
For the MZMs, an 800 nm-thick SiO$_2$ cladding layer is then deposited by plasma-enhanced chemical vapor deposition.
The ring resonators used to evaluate optical loss are left un-cladded.
Trenches for the electrodes are patterned by 375 nm photolithography with SPR700-1.0 resist and are subsequently dry etched using C$_3$F$_8$ and Ar$^{+}$ gases. 
Electron-beam metal evaporation and lift-off is used to define the electrodes (800 nm-thick Au on 15 nm of Ti). 
All MZMs are hotplate-heated at 300°C for 5 h to remove trapped charges which can negatively impact DC drift effects. 
The TFLN MZM is fabricated using a similar process with the same waveguide and electrode geometries.
Further fabrication details are outlined in our previous work \cite{powell2024stable}.

\section{Results}

A scanning electron microscope image of one of the grating couplers is shown in Fig. \ref{fig:figure1}c. 
Using a supercontinuum source and separate chip with grating couplers connected by a waveguide, we estimate 3 dB-bandwidth of the coupler to be 35 nm, with a peak efficiency of a few percent (15.7 dB loss) per coupler.  
Further design and fabrication improvements are expected to increase grating coupling efficiency to 30 per-cent (5 dB loss). 
Next, using continuous-wave laser light at 737 nm, we direct light through the MZM and estimate its loss to be 5.3 dB (excluding grating coupler loss) across a 28 mm device length including routing waveguides. 
The transmission is mainly limited by metal absorption, bending loss, and fabrication-induced sidewall roughness.

Next we characterize the electro-optic performance of the MZM using 737 nm laser light with 4.3 dBm on-chip power.
First, a Hz-rate varied applied voltage reveals a high extinction ratio of 29.6 dB (Fig. \ref{fig:figure1}d).
This ratio suggests the directional couplers have a splitting ratio of 49.5:50.5, which are near optimal of 50:50. 
This measurement yields a V$_{\pi}$ of 1.3 V (Fig. \ref{fig:figure1}d), corresponding to a low 0.65 V$\cdot$cm V$_\pi$L, which is comparable to TFLN near-IR MZMs \cite{renaud2023sub}.
We then use a Vector Network Analyzer (VNA, Agilent E8364B) to send high frequency electrical signals to the modulator. 
The modulator is biased at quadrature and the resultant modulated optical signals are directed to a 20 GHz-bandwidth photodetector. 
The detector is connected to the VNA to form the electric-optic (EO) measurement loop. 
The EO performance (S$_{21}$) and electrical reflection (S$_{11}$) of the MZM is shown in Fig. \ref{fig:figure1}e.
The 3-dB EO roll-off frequency is $\sim$5 GHz, whereas the 3 dB roll-off frequency in terms of half-wave voltage V$_\pi$, that is the 6 dB line in S$_{21}$ Fig. \ref{fig:figure1}e, exceeds the bandwidth of our photodetector.
The rapid roll-off in combination with flat high-frequency response suggests that our device suffers from imperfect impedance matching rather than velocity matching. 
This can be addressed using a thicker bottom oxide layer in conjunction with redesigned, e.g. segmented, electrodes.
The impedance mismatch is also consistent with the strong reflection measured in S$_{11}$ (Fig. \ref{fig:figure1}e).

\begin{figure*}[ht!]
\centering\includegraphics[width=\linewidth]{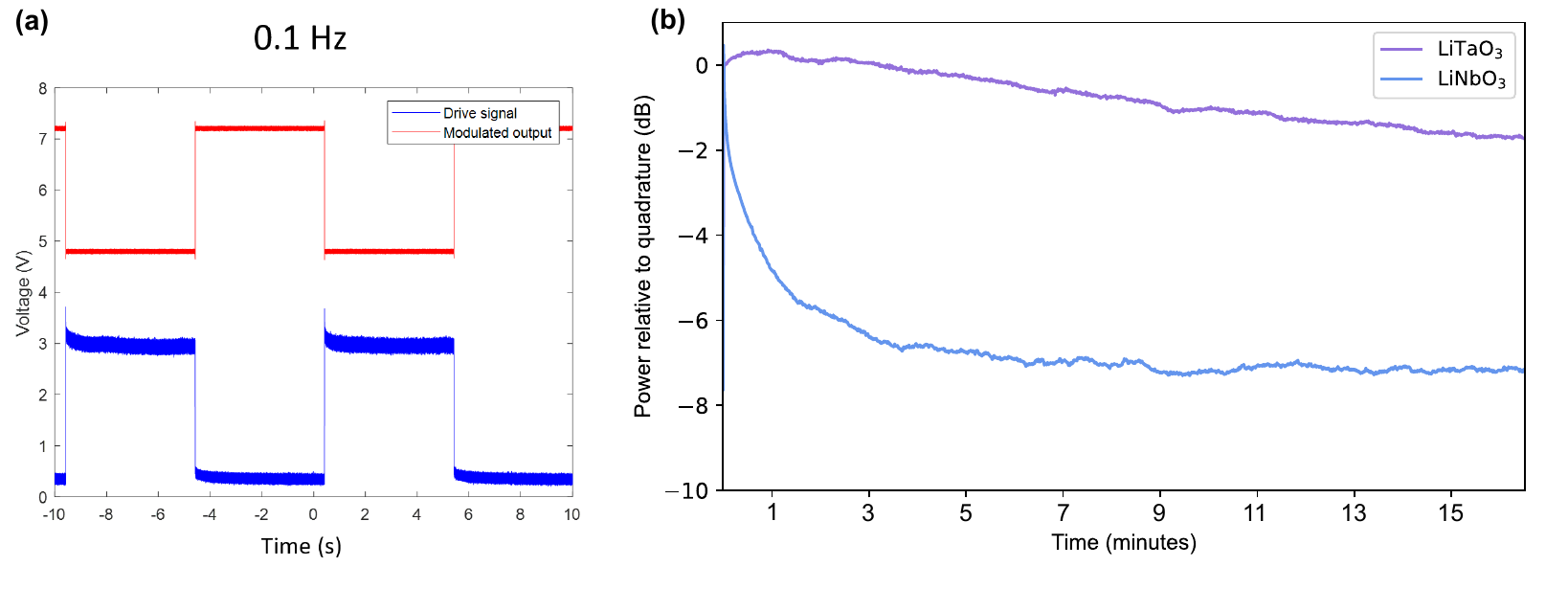}
\caption{Measured electro-optic relaxation of a thin-film lithium tantalate Mach-Zehnder modulator using (a) an applied 0.1 Hz square wave signal and (b) a voltage step.
An equivalent counterpart thin-film lithium niobate modulator relaxes faster under the same conditions.
}
\label{fig:figure2}
\end{figure*}

Next, we measure the DC bias stability of our MZM over long timescales.
First we apply a 0.1 Hz-frequency square wave to the modulator using an on-chip optical power of 4.3 dBm at 737 nm and measure the modulator response with a photodetector.
The input drive signal and corresponding output optical signal is shown in Fig. \ref{fig:figure2}a in red and blue, respectively.
At this frequency electro-optic relaxation is difficult to observe. 
Thus, to elucidate EO relaxation at longer timescales we apply a step voltage (from in-phase to quadrature) to the device and hold the voltage constant.
Using 4.3 dBm of on-chip optical power, we measure a 2 dB of optical power drift over a 16 minute timescale (Fig. \ref{fig:figure2}b). 
We perform the same measurement with 4.3 dBm on-chip power using an equivalent TFLN MZM, which displays a DC bias drift of 8 dB over the same time scale. 

\begin{figure*}[ht!]
\centering\includegraphics[width=\linewidth]{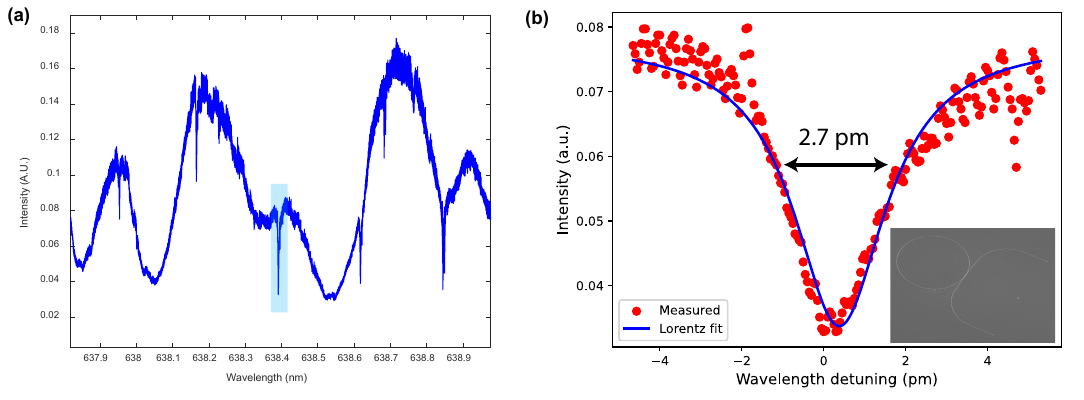}
\caption{(a) Measured transmission spectrum of a thin-film lithium tantalate micro-ring resonator at wavelengths around 638 nm. (b) Resonance linewidth of the micro-ring with Lorentzian fit reveals a FWHM linewidth of 2.7 pm. Inset: scanning electron microwscope image of the fabricated micro-ring resonator and bus waveguide.}
\label{fig:figure3}
\end{figure*}

Finally to further investigate optical propagation loss of our waveguides, we fabricate grating-coupled 239 $\mu$m-diameter micro-ring resonators.
The waveguide width for the rings is 1 $\mu$m.
A SEM image of one of the resonator devices is shown in the inset of Fig. \ref{fig:figure3}b. 
Given the limited tuning range of our 737 nm laser, we instead use a tunable 638 nm-wavelength laser and measure the resonance spectrum of a ring (Fig. \ref{fig:figure3}a).
Laser power fluctuations are observed due to multimode output of the laser and coupling setup (Fig. \ref{fig:figure3}a inset), which we calibrate away to clearly observe the 205 pm free spectral range of our ring (Fig. \ref{fig:figure3}a).
A fit of the resonance linewidth produces a FWHM of 2.7 pm (Fig. \ref{fig:figure3}b), corresponding to a loaded quality (Q) factor of $2.8 \times 10^5$.
Since the resonator is strongly overcoupled, this approximates the intrinsic quality factor, which corresponds to a loss coefficient of 0.5 dB/cm.
We adjusted the laser power to ensure that the resonance linewidth measurement was not impacted by thermo-optic or photo-refractive effects.
This loss coefficient is comparable to that measured in TFLN \cite{renaud2023sub}.

\section{Conclusion}

We demonstrated a thin-film lithium tantalate Mach-Zehnder electro-optic modulator with a slow EO relaxation and low half-wave voltage length product of 0.65 V$\cdot$cm  of interest for applications of near-IR opto-electronics. 
The modulator features a low optical loss of 5.3 dB, an extinction ratio of 30 dB, and a detector-limited bandwidth of 20 GHz, with the latter being suitable for several spectroscopic and atomic applications, including interfacing with the SiV$^-$ center in diamond.
A balanced MZM design, as we have demonstrated at telecommunication wavelength \cite{powell2024stable}, could improve the DC stability further in addition to improved material processing strategies: annealing, doping, surface treatments or different electrode metals.

\begin{acknowledgments}
 The authors thank M. Yeh and D. Barton for discussions.
We acknowledge funding from NSF EEC-1941583, AFOSR FA9550-20-1-01015, NSF 2138068, NASA 80NSSC22K0262, MagiQ Technology/Naval Air Warfare Center N6833522C0413, and Amazon Web Services.
This work was performed in part at the Harvard University Center for Nanoscale Systems (CNS); a member of the National Nanotechnology Coordinated Infrastructure Network (NNCI), which is supported by the National Science Foundation under NSF award no. ECCS-2025158.
\end{acknowledgments}

\section*{Author Declarations} 
\subsection*{Conflict of Interest}K.P., N.S., and M.L. are involved in developing lithium tantalate technologies at Lumina Corporation.
D.R. and M.L. are involved in developing lithium niobate technologies at HyperLight Corporation.

\section*{Data Availability Statement}
Data available on request from the authors.

\nocite{*}
\bibliography{aipsamp}

\begin{thebibliography}{55}%
\makeatletter
\providecommand \@ifxundefined [1]{%
 \@ifx{#1\undefined}
}%
\providecommand \@ifnum [1]{%
 \ifnum #1\expandafter \@firstoftwo
 \else \expandafter \@secondoftwo
 \fi
}%
\providecommand \@ifx [1]{%
 \ifx #1\expandafter \@firstoftwo
 \else \expandafter \@secondoftwo
 \fi
}%
\providecommand \natexlab [1]{#1}%
\providecommand \enquote  [1]{``#1''}%
\providecommand \bibnamefont  [1]{#1}%
\providecommand \bibfnamefont [1]{#1}%
\providecommand \citenamefont [1]{#1}%
\providecommand \href@noop [0]{\@secondoftwo}%
\providecommand \href [0]{\begingroup \@sanitize@url \@href}%
\providecommand \@href[1]{\@@startlink{#1}\@@href}%
\providecommand \@@href[1]{\endgroup#1\@@endlink}%
\providecommand \@sanitize@url [0]{\catcode `\\12\catcode `\$12\catcode `\&12\catcode `\#12\catcode `\^12\catcode `\_12\catcode `\%12\relax}%
\providecommand \@@startlink[1]{}%
\providecommand \@@endlink[0]{}%
\providecommand \url  [0]{\begingroup\@sanitize@url \@url }%
\providecommand \@url [1]{\endgroup\@href {#1}{\urlprefix }}%
\providecommand \urlprefix  [0]{URL }%
\providecommand \Eprint [0]{\href }%
\providecommand \doibase [0]{http://dx.doi.org/}%
\providecommand \selectlanguage [0]{\@gobble}%
\providecommand \bibinfo  [0]{\@secondoftwo}%
\providecommand \bibfield  [0]{\@secondoftwo}%
\providecommand \translation [1]{[#1]}%
\providecommand \BibitemOpen [0]{}%
\providecommand \bibitemStop [0]{}%
\providecommand \bibitemNoStop [0]{.\EOS\space}%
\providecommand \EOS [0]{\spacefactor3000\relax}%
\providecommand \BibitemShut  [1]{\csname bibitem#1\endcsname}%
\let\auto@bib@innerbib\@empty
\bibitem [{\citenamefont {Hu}\ \emph {et~al.}(2025)\citenamefont {Hu}, \citenamefont {Zhu}, \citenamefont {Lu}, \citenamefont {Zhu}, \citenamefont {Song}, \citenamefont {Renaud}, \citenamefont {Assumpcao}, \citenamefont {Cheng}, \citenamefont {Xin}, \citenamefont {Yeh} \emph {et~al.}}]{hu2025integrated}%
  \BibitemOpen
  \bibfield  {author} {\bibinfo {author} {\bibfnamefont {Y.}~\bibnamefont {Hu}}, \bibinfo {author} {\bibfnamefont {D.}~\bibnamefont {Zhu}}, \bibinfo {author} {\bibfnamefont {S.}~\bibnamefont {Lu}}, \bibinfo {author} {\bibfnamefont {X.}~\bibnamefont {Zhu}}, \bibinfo {author} {\bibfnamefont {Y.}~\bibnamefont {Song}}, \bibinfo {author} {\bibfnamefont {D.}~\bibnamefont {Renaud}}, \bibinfo {author} {\bibfnamefont {D.}~\bibnamefont {Assumpcao}}, \bibinfo {author} {\bibfnamefont {R.}~\bibnamefont {Cheng}}, \bibinfo {author} {\bibfnamefont {C.}~\bibnamefont {Xin}}, \bibinfo {author} {\bibfnamefont {M.}~\bibnamefont {Yeh}},  \emph {et~al.},\ }\bibfield  {title} {\enquote {\bibinfo {title} {Integrated electro-optics on thin-film lithium niobate},}\ }\href@noop {} {\bibfield  {journal} {\bibinfo  {journal} {Nature Reviews Physics}\ ,\ \bibinfo {pages} {1--18}} (\bibinfo {year} {2025})}\BibitemShut {NoStop}%
\bibitem [{\citenamefont {Boes}\ \emph {et~al.}(2023)\citenamefont {Boes}, \citenamefont {Chang}, \citenamefont {Langrock}, \citenamefont {Yu}, \citenamefont {Zhang}, \citenamefont {Lin}, \citenamefont {Lon{\v{c}}ar}, \citenamefont {Fejer}, \citenamefont {Bowers},\ and\ \citenamefont {Mitchell}}]{boes2023lithium}%
  \BibitemOpen
  \bibfield  {author} {\bibinfo {author} {\bibfnamefont {A.}~\bibnamefont {Boes}}, \bibinfo {author} {\bibfnamefont {L.}~\bibnamefont {Chang}}, \bibinfo {author} {\bibfnamefont {C.}~\bibnamefont {Langrock}}, \bibinfo {author} {\bibfnamefont {M.}~\bibnamefont {Yu}}, \bibinfo {author} {\bibfnamefont {M.}~\bibnamefont {Zhang}}, \bibinfo {author} {\bibfnamefont {Q.}~\bibnamefont {Lin}}, \bibinfo {author} {\bibfnamefont {M.}~\bibnamefont {Lon{\v{c}}ar}}, \bibinfo {author} {\bibfnamefont {M.}~\bibnamefont {Fejer}}, \bibinfo {author} {\bibfnamefont {J.}~\bibnamefont {Bowers}}, \ and\ \bibinfo {author} {\bibfnamefont {A.}~\bibnamefont {Mitchell}},\ }\bibfield  {title} {\enquote {\bibinfo {title} {Lithium niobate photonics: Unlocking the electromagnetic spectrum},}\ }\href@noop {} {\bibfield  {journal} {\bibinfo  {journal} {Science}\ }\textbf {\bibinfo {volume} {379}} (\bibinfo {year} {2023})}\BibitemShut {NoStop}%
\bibitem [{\citenamefont {Zhu}\ \emph {et~al.}(2021{\natexlab{a}})\citenamefont {Zhu}, \citenamefont {Shao}, \citenamefont {Yu}, \citenamefont {Cheng}, \citenamefont {Desiatov}, \citenamefont {Xin}, \citenamefont {Hu}, \citenamefont {Holzgrafe}, \citenamefont {Ghosh}, \citenamefont {Shams-Ansari} \emph {et~al.}}]{zhu2021integrated}%
  \BibitemOpen
  \bibfield  {author} {\bibinfo {author} {\bibfnamefont {D.}~\bibnamefont {Zhu}}, \bibinfo {author} {\bibfnamefont {L.}~\bibnamefont {Shao}}, \bibinfo {author} {\bibfnamefont {M.}~\bibnamefont {Yu}}, \bibinfo {author} {\bibfnamefont {R.}~\bibnamefont {Cheng}}, \bibinfo {author} {\bibfnamefont {B.}~\bibnamefont {Desiatov}}, \bibinfo {author} {\bibfnamefont {C.}~\bibnamefont {Xin}}, \bibinfo {author} {\bibfnamefont {Y.}~\bibnamefont {Hu}}, \bibinfo {author} {\bibfnamefont {J.}~\bibnamefont {Holzgrafe}}, \bibinfo {author} {\bibfnamefont {S.}~\bibnamefont {Ghosh}}, \bibinfo {author} {\bibfnamefont {A.}~\bibnamefont {Shams-Ansari}},  \emph {et~al.},\ }\bibfield  {title} {\enquote {\bibinfo {title} {Integrated photonics on thin-film lithium niobate},}\ }\href@noop {} {\bibfield  {journal} {\bibinfo  {journal} {Advances in Optics and Photonics}\ }\textbf {\bibinfo {volume} {13}},\ \bibinfo {pages} {242--352} (\bibinfo {year} {2021}{\natexlab{a}})}\BibitemShut {NoStop}%
\bibitem [{\citenamefont {Xu}\ \emph {et~al.}(2020)\citenamefont {Xu}, \citenamefont {He}, \citenamefont {Zhang}, \citenamefont {Jian}, \citenamefont {Pan}, \citenamefont {Liu}, \citenamefont {Chen}, \citenamefont {Meng}, \citenamefont {Chen}, \citenamefont {Li} \emph {et~al.}}]{xu2020high}%
  \BibitemOpen
  \bibfield  {author} {\bibinfo {author} {\bibfnamefont {M.}~\bibnamefont {Xu}}, \bibinfo {author} {\bibfnamefont {M.}~\bibnamefont {He}}, \bibinfo {author} {\bibfnamefont {H.}~\bibnamefont {Zhang}}, \bibinfo {author} {\bibfnamefont {J.}~\bibnamefont {Jian}}, \bibinfo {author} {\bibfnamefont {Y.}~\bibnamefont {Pan}}, \bibinfo {author} {\bibfnamefont {X.}~\bibnamefont {Liu}}, \bibinfo {author} {\bibfnamefont {L.}~\bibnamefont {Chen}}, \bibinfo {author} {\bibfnamefont {X.}~\bibnamefont {Meng}}, \bibinfo {author} {\bibfnamefont {H.}~\bibnamefont {Chen}}, \bibinfo {author} {\bibfnamefont {Z.}~\bibnamefont {Li}},  \emph {et~al.},\ }\bibfield  {title} {\enquote {\bibinfo {title} {High-performance coherent optical modulators based on thin-film lithium niobate platform},}\ }\href@noop {} {\bibfield  {journal} {\bibinfo  {journal} {Nature communications}\ }\textbf {\bibinfo {volume} {11}},\ \bibinfo {pages} {3911} (\bibinfo {year} {2020})}\BibitemShut {NoStop}%
\bibitem [{\citenamefont {Desiatov}\ \emph {et~al.}(2019)\citenamefont {Desiatov}, \citenamefont {Shams-Ansari}, \citenamefont {Zhang}, \citenamefont {Wang},\ and\ \citenamefont {Lon{\v{c}}ar}}]{desiatov2019ultra}%
  \BibitemOpen
  \bibfield  {author} {\bibinfo {author} {\bibfnamefont {B.}~\bibnamefont {Desiatov}}, \bibinfo {author} {\bibfnamefont {A.}~\bibnamefont {Shams-Ansari}}, \bibinfo {author} {\bibfnamefont {M.}~\bibnamefont {Zhang}}, \bibinfo {author} {\bibfnamefont {C.}~\bibnamefont {Wang}}, \ and\ \bibinfo {author} {\bibfnamefont {M.}~\bibnamefont {Lon{\v{c}}ar}},\ }\bibfield  {title} {\enquote {\bibinfo {title} {Ultra-low-loss integrated visible photonics using thin-film lithium niobate},}\ }\href@noop {} {\bibfield  {journal} {\bibinfo  {journal} {Optica}\ }\textbf {\bibinfo {volume} {6}},\ \bibinfo {pages} {380--384} (\bibinfo {year} {2019})}\BibitemShut {NoStop}%
\bibitem [{\citenamefont {Powell}\ \emph {et~al.}(2024)\citenamefont {Powell}, \citenamefont {Li}, \citenamefont {Assumpcao}, \citenamefont {Magalh{\~a}es}, \citenamefont {Sinclair},\ and\ \citenamefont {Lon{\v{c}}ar}}]{powell2024stable}%
  \BibitemOpen
  \bibfield  {author} {\bibinfo {author} {\bibfnamefont {K.}~\bibnamefont {Powell}}, \bibinfo {author} {\bibfnamefont {X.}~\bibnamefont {Li}}, \bibinfo {author} {\bibfnamefont {D.}~\bibnamefont {Assumpcao}}, \bibinfo {author} {\bibfnamefont {L.}~\bibnamefont {Magalh{\~a}es}}, \bibinfo {author} {\bibfnamefont {N.}~\bibnamefont {Sinclair}}, \ and\ \bibinfo {author} {\bibfnamefont {M.}~\bibnamefont {Lon{\v{c}}ar}},\ }\bibfield  {title} {\enquote {\bibinfo {title} {Dc-stable electro-optic modulators using thin-film lithium tantalate},}\ }\href@noop {} {\bibfield  {journal} {\bibinfo  {journal} {Optics Express}\ }\textbf {\bibinfo {volume} {32}},\ \bibinfo {pages} {44115--44122} (\bibinfo {year} {2024})}\BibitemShut {NoStop}%
\bibitem [{\citenamefont {Wang}\ \emph {et~al.}(2024{\natexlab{a}})\citenamefont {Wang}, \citenamefont {Li}, \citenamefont {Riemensberger}, \citenamefont {Lihachev}, \citenamefont {Churaev}, \citenamefont {Kao}, \citenamefont {Ji}, \citenamefont {Zhang}, \citenamefont {Blesin}, \citenamefont {Davydova}, \citenamefont {Chen}, \citenamefont {Huang}, \citenamefont {Wang}, \citenamefont {Ou},\ and\ \citenamefont {Kippenberg}}]{wang2023}%
  \BibitemOpen
  \bibfield  {author} {\bibinfo {author} {\bibfnamefont {C.}~\bibnamefont {Wang}}, \bibinfo {author} {\bibfnamefont {Z.}~\bibnamefont {Li}}, \bibinfo {author} {\bibfnamefont {J.}~\bibnamefont {Riemensberger}}, \bibinfo {author} {\bibfnamefont {G.}~\bibnamefont {Lihachev}}, \bibinfo {author} {\bibfnamefont {M.}~\bibnamefont {Churaev}}, \bibinfo {author} {\bibfnamefont {W.}~\bibnamefont {Kao}}, \bibinfo {author} {\bibfnamefont {X.}~\bibnamefont {Ji}}, \bibinfo {author} {\bibfnamefont {J.}~\bibnamefont {Zhang}}, \bibinfo {author} {\bibfnamefont {T.}~\bibnamefont {Blesin}}, \bibinfo {author} {\bibfnamefont {A.}~\bibnamefont {Davydova}}, \bibinfo {author} {\bibfnamefont {Y.}~\bibnamefont {Chen}}, \bibinfo {author} {\bibfnamefont {K.}~\bibnamefont {Huang}}, \bibinfo {author} {\bibfnamefont {X.}~\bibnamefont {Wang}}, \bibinfo {author} {\bibfnamefont {X.}~\bibnamefont {Ou}}, \ and\ \bibinfo {author} {\bibfnamefont {T.~J.}\ \bibnamefont {Kippenberg}},\ }\bibfield  {title} {\enquote {\bibinfo {title} {Lithium tantalate
  photonic integrated circuits for volume manufacturing},}\ }\href {\doibase 10.1038/s41586-024-07369-1} {\bibfield  {journal} {\bibinfo  {journal} {Nature}\ }\textbf {\bibinfo {volume} {629}},\ \bibinfo {pages} {784--790} (\bibinfo {year} {2024}{\natexlab{a}})}\BibitemShut {NoStop}%
\bibitem [{\citenamefont {Yu}\ \emph {et~al.}(2024{\natexlab{a}})\citenamefont {Yu}, \citenamefont {Ruan}, \citenamefont {Xue}, \citenamefont {Wang}, \citenamefont {Gan}, \citenamefont {Gao}, \citenamefont {Guo}, \citenamefont {Chen}, \citenamefont {Ou},\ and\ \citenamefont {Liu}}]{DC_EO_LT}%
  \BibitemOpen
  \bibfield  {author} {\bibinfo {author} {\bibfnamefont {J.}~\bibnamefont {Yu}}, \bibinfo {author} {\bibfnamefont {Z.}~\bibnamefont {Ruan}}, \bibinfo {author} {\bibfnamefont {Y.}~\bibnamefont {Xue}}, \bibinfo {author} {\bibfnamefont {H.}~\bibnamefont {Wang}}, \bibinfo {author} {\bibfnamefont {R.}~\bibnamefont {Gan}}, \bibinfo {author} {\bibfnamefont {T.}~\bibnamefont {Gao}}, \bibinfo {author} {\bibfnamefont {C.}~\bibnamefont {Guo}}, \bibinfo {author} {\bibfnamefont {K.}~\bibnamefont {Chen}}, \bibinfo {author} {\bibfnamefont {X.}~\bibnamefont {Ou}}, \ and\ \bibinfo {author} {\bibfnamefont {L.}~\bibnamefont {Liu}},\ }\bibfield  {title} {\enquote {\bibinfo {title} {{Tunable and stable micro-ring resonator based on thin-film lithium tantalate}},}\ }\href {\doibase 10.1063/5.0187996} {\bibfield  {journal} {\bibinfo  {journal} {APL Photonics}\ }\textbf {\bibinfo {volume} {9}},\ \bibinfo {pages} {036115} (\bibinfo {year} {2024}{\natexlab{a}})},\ \Eprint
  {http://arxiv.org/abs/https://pubs.aip.org/aip/app/article-pdf/doi/10.1063/5.0187996/19844291/036115\_1\_5.0187996.pdf} {https://pubs.aip.org/aip/app/article-pdf/doi/10.1063/5.0187996/19844291/036115\_1\_5.0187996.pdf} \BibitemShut {NoStop}%
\bibitem [{\citenamefont {Shen}\ \emph {et~al.}(2023)\citenamefont {Shen}, \citenamefont {Zhang}, \citenamefont {Feng}, \citenamefont {Xu}, \citenamefont {Zhang},\ and\ \citenamefont {Su}}]{shen23}%
  \BibitemOpen
  \bibfield  {author} {\bibinfo {author} {\bibfnamefont {J.}~\bibnamefont {Shen}}, \bibinfo {author} {\bibfnamefont {Y.}~\bibnamefont {Zhang}}, \bibinfo {author} {\bibfnamefont {C.}~\bibnamefont {Feng}}, \bibinfo {author} {\bibfnamefont {Z.}~\bibnamefont {Xu}}, \bibinfo {author} {\bibfnamefont {L.}~\bibnamefont {Zhang}}, \ and\ \bibinfo {author} {\bibfnamefont {Y.}~\bibnamefont {Su}},\ }\bibfield  {title} {\enquote {\bibinfo {title} {Hybrid lithium tantalite-silicon integrated photonics platform for electro-optic modulation},}\ }\href {\doibase 10.1364/OL.502492} {\bibfield  {journal} {\bibinfo  {journal} {Opt. Lett.}\ }\textbf {\bibinfo {volume} {48}},\ \bibinfo {pages} {6176--6179} (\bibinfo {year} {2023})}\BibitemShut {NoStop}%
\bibitem [{\citenamefont {Yu}\ \emph {et~al.}(2024{\natexlab{b}})\citenamefont {Yu}, \citenamefont {Ruan}, \citenamefont {Xue}, \citenamefont {Wang}, \citenamefont {Gan}, \citenamefont {Gao}, \citenamefont {Guo}, \citenamefont {Chen}, \citenamefont {Ou},\ and\ \citenamefont {Liu}}]{yu_24}%
  \BibitemOpen
  \bibfield  {author} {\bibinfo {author} {\bibfnamefont {J.}~\bibnamefont {Yu}}, \bibinfo {author} {\bibfnamefont {Z.}~\bibnamefont {Ruan}}, \bibinfo {author} {\bibfnamefont {Y.}~\bibnamefont {Xue}}, \bibinfo {author} {\bibfnamefont {H.}~\bibnamefont {Wang}}, \bibinfo {author} {\bibfnamefont {R.}~\bibnamefont {Gan}}, \bibinfo {author} {\bibfnamefont {T.}~\bibnamefont {Gao}}, \bibinfo {author} {\bibfnamefont {C.}~\bibnamefont {Guo}}, \bibinfo {author} {\bibfnamefont {K.}~\bibnamefont {Chen}}, \bibinfo {author} {\bibfnamefont {X.}~\bibnamefont {Ou}}, \ and\ \bibinfo {author} {\bibfnamefont {L.}~\bibnamefont {Liu}},\ }\bibfield  {title} {\enquote {\bibinfo {title} {{Tunable and stable micro-ring resonator based on thin-film lithium tantalate}},}\ }\href {\doibase 10.1063/5.0187996} {\bibfield  {journal} {\bibinfo  {journal} {APL Photonics}\ }\textbf {\bibinfo {volume} {9}},\ \bibinfo {pages} {036115} (\bibinfo {year} {2024}{\natexlab{b}})},\ \Eprint
  {http://arxiv.org/abs/https://pubs.aip.org/aip/app/article-pdf/doi/10.1063/5.0187996/19844291/036115\_1\_5.0187996.pdf} {https://pubs.aip.org/aip/app/article-pdf/doi/10.1063/5.0187996/19844291/036115\_1\_5.0187996.pdf} \BibitemShut {NoStop}%
\bibitem [{\citenamefont {Nishi}\ \emph {et~al.}(2022)\citenamefont {Nishi}, \citenamefont {Tsuchizawa}, \citenamefont {Segawa},\ and\ \citenamefont {Matsuo}}]{nishi_22}%
  \BibitemOpen
  \bibfield  {author} {\bibinfo {author} {\bibfnamefont {H.}~\bibnamefont {Nishi}}, \bibinfo {author} {\bibfnamefont {T.}~\bibnamefont {Tsuchizawa}}, \bibinfo {author} {\bibfnamefont {T.}~\bibnamefont {Segawa}}, \ and\ \bibinfo {author} {\bibfnamefont {S.}~\bibnamefont {Matsuo}},\ }\bibfield  {title} {\enquote {\bibinfo {title} {Low-loss lithium tantalate on insulator waveguide towards on-chip nonlinear photonics},}\ }in\ \href {\doibase 10.23919/OECC/PSC53152.2022.9850122} {\emph {\bibinfo {booktitle} {2022 27th OptoElectronics and Communications Conference (OECC) and 2022 International Conference on Photonics in Switching and Computing (PSC)}}}\ (\bibinfo {year} {2022})\ pp.\ \bibinfo {pages} {1--3}\BibitemShut {NoStop}%
\bibitem [{\citenamefont {Wang}\ \emph {et~al.}(2024{\natexlab{b}})\citenamefont {Wang}, \citenamefont {Fang}, \citenamefont {Zhang}, \citenamefont {Kotz}, \citenamefont {Lihachev}, \citenamefont {Churaev}, \citenamefont {Li}, \citenamefont {Schwarzenberger}, \citenamefont {Ou}, \citenamefont {Koos} \emph {et~al.}}]{wang2024ultrabroadband}%
  \BibitemOpen
  \bibfield  {author} {\bibinfo {author} {\bibfnamefont {C.}~\bibnamefont {Wang}}, \bibinfo {author} {\bibfnamefont {D.}~\bibnamefont {Fang}}, \bibinfo {author} {\bibfnamefont {J.}~\bibnamefont {Zhang}}, \bibinfo {author} {\bibfnamefont {A.}~\bibnamefont {Kotz}}, \bibinfo {author} {\bibfnamefont {G.}~\bibnamefont {Lihachev}}, \bibinfo {author} {\bibfnamefont {M.}~\bibnamefont {Churaev}}, \bibinfo {author} {\bibfnamefont {Z.}~\bibnamefont {Li}}, \bibinfo {author} {\bibfnamefont {A.}~\bibnamefont {Schwarzenberger}}, \bibinfo {author} {\bibfnamefont {X.}~\bibnamefont {Ou}}, \bibinfo {author} {\bibfnamefont {C.}~\bibnamefont {Koos}},  \emph {et~al.},\ }\bibfield  {title} {\enquote {\bibinfo {title} {Ultrabroadband thin-film lithium tantalate modulator for high-speed communications},}\ }\href@noop {} {\bibfield  {journal} {\bibinfo  {journal} {Optica}\ }\textbf {\bibinfo {volume} {11}},\ \bibinfo {pages} {1614--1620} (\bibinfo {year} {2024}{\natexlab{b}})}\BibitemShut {NoStop}%
\bibitem [{\citenamefont {Zhang}\ \emph {et~al.}(2025)\citenamefont {Zhang}, \citenamefont {Wang}, \citenamefont {Denney}, \citenamefont {Riemensberger}, \citenamefont {Lihachev}, \citenamefont {Hu}, \citenamefont {Kao}, \citenamefont {Bl{\'e}sin}, \citenamefont {Kuznetsov}, \citenamefont {Li} \emph {et~al.}}]{zhang2025ultrabroadband}%
  \BibitemOpen
  \bibfield  {author} {\bibinfo {author} {\bibfnamefont {J.}~\bibnamefont {Zhang}}, \bibinfo {author} {\bibfnamefont {C.}~\bibnamefont {Wang}}, \bibinfo {author} {\bibfnamefont {C.}~\bibnamefont {Denney}}, \bibinfo {author} {\bibfnamefont {J.}~\bibnamefont {Riemensberger}}, \bibinfo {author} {\bibfnamefont {G.}~\bibnamefont {Lihachev}}, \bibinfo {author} {\bibfnamefont {J.}~\bibnamefont {Hu}}, \bibinfo {author} {\bibfnamefont {W.}~\bibnamefont {Kao}}, \bibinfo {author} {\bibfnamefont {T.}~\bibnamefont {Bl{\'e}sin}}, \bibinfo {author} {\bibfnamefont {N.}~\bibnamefont {Kuznetsov}}, \bibinfo {author} {\bibfnamefont {Z.}~\bibnamefont {Li}},  \emph {et~al.},\ }\bibfield  {title} {\enquote {\bibinfo {title} {Ultrabroadband integrated electro-optic frequency comb in lithium tantalate},}\ }\href@noop {} {\bibfield  {journal} {\bibinfo  {journal} {Nature}\ ,\ \bibinfo {pages} {1--8}} (\bibinfo {year} {2025})}\BibitemShut {NoStop}%
\bibitem [{\citenamefont {Casson}\ \emph {et~al.}(2004)\citenamefont {Casson}, \citenamefont {Gahagan}, \citenamefont {Scrymgeour}, \citenamefont {Jain}, \citenamefont {Robinson}, \citenamefont {Gopalan},\ and\ \citenamefont {Sander}}]{LT_pockels}%
  \BibitemOpen
  \bibfield  {author} {\bibinfo {author} {\bibfnamefont {J.~L.}\ \bibnamefont {Casson}}, \bibinfo {author} {\bibfnamefont {K.~T.}\ \bibnamefont {Gahagan}}, \bibinfo {author} {\bibfnamefont {D.~A.}\ \bibnamefont {Scrymgeour}}, \bibinfo {author} {\bibfnamefont {R.~K.}\ \bibnamefont {Jain}}, \bibinfo {author} {\bibfnamefont {J.~M.}\ \bibnamefont {Robinson}}, \bibinfo {author} {\bibfnamefont {V.}~\bibnamefont {Gopalan}}, \ and\ \bibinfo {author} {\bibfnamefont {R.~K.}\ \bibnamefont {Sander}},\ }\bibfield  {title} {\enquote {\bibinfo {title} {Electro-optic coefficients of lithium tantalate at near-infrared wavelengths},}\ }\href {\doibase 10.1364/JOSAB.21.001948} {\bibfield  {journal} {\bibinfo  {journal} {J. Opt. Soc. Am. B}\ }\textbf {\bibinfo {volume} {21}},\ \bibinfo {pages} {1948--1952} (\bibinfo {year} {2004})}\BibitemShut {NoStop}%
\bibitem [{\citenamefont {Çabuk}\ and\ \citenamefont {Mamedov}(1999)}]{LT_bandgap}%
  \BibitemOpen
  \bibfield  {author} {\bibinfo {author} {\bibfnamefont {S.}~\bibnamefont {Çabuk}}\ and\ \bibinfo {author} {\bibfnamefont {A.}~\bibnamefont {Mamedov}},\ }\bibfield  {title} {\enquote {\bibinfo {title} {Urbach rule and optical properties of the linbo$_3$ and litao$_3$},}\ }\href {\doibase 10.1088/1464-4258/1/3/313} {\bibfield  {journal} {\bibinfo  {journal} {Journal of Optics A: Pure and Applied Optics}\ }\textbf {\bibinfo {volume} {1}},\ \bibinfo {pages} {424} (\bibinfo {year} {1999})}\BibitemShut {NoStop}%
\bibitem [{\citenamefont {Jacob}\ \emph {et~al.}(2004)\citenamefont {Jacob}, \citenamefont {Hartnett}, \citenamefont {Mazierska}, \citenamefont {Giordano}, \citenamefont {Krupka},\ and\ \citenamefont {Tobar}}]{LT_good}%
  \BibitemOpen
  \bibfield  {author} {\bibinfo {author} {\bibfnamefont {M.}~\bibnamefont {Jacob}}, \bibinfo {author} {\bibfnamefont {J.}~\bibnamefont {Hartnett}}, \bibinfo {author} {\bibfnamefont {J.}~\bibnamefont {Mazierska}}, \bibinfo {author} {\bibfnamefont {V.}~\bibnamefont {Giordano}}, \bibinfo {author} {\bibfnamefont {J.}~\bibnamefont {Krupka}}, \ and\ \bibinfo {author} {\bibfnamefont {M.}~\bibnamefont {Tobar}},\ }\bibfield  {title} {\enquote {\bibinfo {title} {Temperature dependence of permittivity and loss tangent of lithium tantalate at microwave frequencies},}\ }\href {\doibase 10.1109/TMTT.2003.821911} {\bibfield  {journal} {\bibinfo  {journal} {IEEE Transactions on Microwave Theory and Techniques}\ }\textbf {\bibinfo {volume} {52}},\ \bibinfo {pages} {536--541} (\bibinfo {year} {2004})}\BibitemShut {NoStop}%
\bibitem [{\citenamefont {Holtmann}\ \emph {et~al.}(2004)\citenamefont {Holtmann}, \citenamefont {Imbrock}, \citenamefont {Bäumer}, \citenamefont {Hesse}, \citenamefont {Krätzig},\ and\ \citenamefont {Kip}}]{LT_phr}%
  \BibitemOpen
  \bibfield  {author} {\bibinfo {author} {\bibfnamefont {F.}~\bibnamefont {Holtmann}}, \bibinfo {author} {\bibfnamefont {J.}~\bibnamefont {Imbrock}}, \bibinfo {author} {\bibfnamefont {C.}~\bibnamefont {Bäumer}}, \bibinfo {author} {\bibfnamefont {H.}~\bibnamefont {Hesse}}, \bibinfo {author} {\bibfnamefont {E.}~\bibnamefont {Krätzig}}, \ and\ \bibinfo {author} {\bibfnamefont {D.}~\bibnamefont {Kip}},\ }\bibfield  {title} {\enquote {\bibinfo {title} {{Photorefractive properties of undoped lithium tantalate crystals for various composition}},}\ }\href {\doibase 10.1063/1.1805189} {\bibfield  {journal} {\bibinfo  {journal} {Journal of Applied Physics}\ }\textbf {\bibinfo {volume} {96}},\ \bibinfo {pages} {7455--7459} (\bibinfo {year} {2004})},\ \Eprint {http://arxiv.org/abs/https://pubs.aip.org/aip/jap/article-pdf/96/12/7455/18720753/7455\_1\_online.pdf} {https://pubs.aip.org/aip/jap/article-pdf/96/12/7455/18720753/7455\_1\_online.pdf} \BibitemShut {NoStop}%
\bibitem [{\citenamefont {Althoff}\ and\ \citenamefont {Kraetzig}(1990)}]{LN_phr}%
  \BibitemOpen
  \bibfield  {author} {\bibinfo {author} {\bibfnamefont {O.}~\bibnamefont {Althoff}}\ and\ \bibinfo {author} {\bibfnamefont {E.~E.}\ \bibnamefont {Kraetzig}},\ }\bibfield  {title} {\enquote {\bibinfo {title} {{Strong light-induced refractive index changes in LiNbO$3$}},}\ }in\ \href {\doibase 10.1117/12.20458} {\emph {\bibinfo {booktitle} {Nonlinear Optical Materials III}}},\ Vol.\ \bibinfo {volume} {1273},\ \bibinfo {editor} {edited by\ \bibinfo {editor} {\bibfnamefont {P.}~\bibnamefont {Guenter}}},\ \bibinfo {organization} {International Society for Optics and Photonics}\ (\bibinfo  {publisher} {SPIE},\ \bibinfo {year} {1990})\ pp.\ \bibinfo {pages} {12 -- 19}\BibitemShut {NoStop}%
\bibitem [{\citenamefont {Yan}\ \emph {et~al.}(2020)\citenamefont {Yan}, \citenamefont {Liu}, \citenamefont {Ge}, \citenamefont {Zhu}, \citenamefont {Wu}, \citenamefont {Chen},\ and\ \citenamefont {Chen}}]{LT_damage}%
  \BibitemOpen
  \bibfield  {author} {\bibinfo {author} {\bibfnamefont {X.}~\bibnamefont {Yan}}, \bibinfo {author} {\bibfnamefont {Y.}~\bibnamefont {Liu}}, \bibinfo {author} {\bibfnamefont {L.}~\bibnamefont {Ge}}, \bibinfo {author} {\bibfnamefont {B.}~\bibnamefont {Zhu}}, \bibinfo {author} {\bibfnamefont {J.}~\bibnamefont {Wu}}, \bibinfo {author} {\bibfnamefont {Y.}~\bibnamefont {Chen}}, \ and\ \bibinfo {author} {\bibfnamefont {X.}~\bibnamefont {Chen}},\ }\bibfield  {title} {\enquote {\bibinfo {title} {High optical damage threshold on-chip lithium tantalate microdisk resonator},}\ }\href {\doibase 10.1364/OL.394171} {\bibfield  {journal} {\bibinfo  {journal} {Opt. Lett.}\ }\textbf {\bibinfo {volume} {45}},\ \bibinfo {pages} {4100--4103} (\bibinfo {year} {2020})}\BibitemShut {NoStop}%
\bibitem [{\citenamefont {Kong}\ \emph {et~al.}(2020)\citenamefont {Kong}, \citenamefont {Bo}, \citenamefont {Wang}, \citenamefont {Zheng}, \citenamefont {Liu}, \citenamefont {Zhang}, \citenamefont {Rupp},\ and\ \citenamefont {Xu}}]{LN_damage}%
  \BibitemOpen
  \bibfield  {author} {\bibinfo {author} {\bibfnamefont {Y.}~\bibnamefont {Kong}}, \bibinfo {author} {\bibfnamefont {F.}~\bibnamefont {Bo}}, \bibinfo {author} {\bibfnamefont {W.}~\bibnamefont {Wang}}, \bibinfo {author} {\bibfnamefont {D.}~\bibnamefont {Zheng}}, \bibinfo {author} {\bibfnamefont {H.}~\bibnamefont {Liu}}, \bibinfo {author} {\bibfnamefont {G.}~\bibnamefont {Zhang}}, \bibinfo {author} {\bibfnamefont {R.}~\bibnamefont {Rupp}}, \ and\ \bibinfo {author} {\bibfnamefont {J.}~\bibnamefont {Xu}},\ }\bibfield  {title} {\enquote {\bibinfo {title} {Recent progress in lithium niobate: Optical damage, defect simulation, and on-chip devices},}\ }\href {\doibase https://doi.org/10.1002/adma.201806452} {\bibfield  {journal} {\bibinfo  {journal} {Advanced Materials}\ }\textbf {\bibinfo {volume} {32}},\ \bibinfo {pages} {1806452} (\bibinfo {year} {2020})},\ \Eprint {http://arxiv.org/abs/https://onlinelibrary.wiley.com/doi/pdf/10.1002/adma.201806452} {https://onlinelibrary.wiley.com/doi/pdf/10.1002/adma.201806452}
  \BibitemShut {NoStop}%
\bibitem [{\citenamefont {Yang}\ \emph {et~al.}(2007)\citenamefont {Yang}, \citenamefont {Su}, \citenamefont {Weng}, \citenamefont {Hung},\ and\ \citenamefont {Wu}}]{LN_losstan}%
  \BibitemOpen
  \bibfield  {author} {\bibinfo {author} {\bibfnamefont {R.-Y.}\ \bibnamefont {Yang}}, \bibinfo {author} {\bibfnamefont {Y.-K.}\ \bibnamefont {Su}}, \bibinfo {author} {\bibfnamefont {M.-H.}\ \bibnamefont {Weng}}, \bibinfo {author} {\bibfnamefont {C.-Y.}\ \bibnamefont {Hung}}, \ and\ \bibinfo {author} {\bibfnamefont {H.-W.}\ \bibnamefont {Wu}},\ }\bibfield  {title} {\enquote {\bibinfo {title} {{Characteristics of coplanar waveguide on lithium niobate crystals as a microwave substrate}},}\ }\href {\doibase 10.1063/1.2402978} {\bibfield  {journal} {\bibinfo  {journal} {Journal of Applied Physics}\ }\textbf {\bibinfo {volume} {101}},\ \bibinfo {pages} {014101} (\bibinfo {year} {2007})},\ \Eprint {http://arxiv.org/abs/https://pubs.aip.org/aip/jap/article-pdf/doi/10.1063/1.2402978/13341406/014101\_1\_online.pdf} {https://pubs.aip.org/aip/jap/article-pdf/doi/10.1063/1.2402978/13341406/014101\_1\_online.pdf} \BibitemShut {NoStop}%
\bibitem [{\citenamefont {Holzgrafe}\ \emph {et~al.}(2024)\citenamefont {Holzgrafe}, \citenamefont {Puma}, \citenamefont {Cheng}, \citenamefont {Warner}, \citenamefont {Shams-Ansari}, \citenamefont {Shankar},\ and\ \citenamefont {Lon\v{c}ar}}]{holzgrafe_TFLN_relaxation}%
  \BibitemOpen
  \bibfield  {author} {\bibinfo {author} {\bibfnamefont {J.}~\bibnamefont {Holzgrafe}}, \bibinfo {author} {\bibfnamefont {E.}~\bibnamefont {Puma}}, \bibinfo {author} {\bibfnamefont {R.}~\bibnamefont {Cheng}}, \bibinfo {author} {\bibfnamefont {H.}~\bibnamefont {Warner}}, \bibinfo {author} {\bibfnamefont {A.}~\bibnamefont {Shams-Ansari}}, \bibinfo {author} {\bibfnamefont {R.}~\bibnamefont {Shankar}}, \ and\ \bibinfo {author} {\bibfnamefont {M.}~\bibnamefont {Lon\v{c}ar}},\ }\bibfield  {title} {\enquote {\bibinfo {title} {Relaxation of the electro-optic response in thin-film lithium niobate modulators},}\ }\href {\doibase 10.1364/OE.507536} {\bibfield  {journal} {\bibinfo  {journal} {Opt. Express}\ }\textbf {\bibinfo {volume} {32}},\ \bibinfo {pages} {3619--3631} (\bibinfo {year} {2024})}\BibitemShut {NoStop}%
\bibitem [{\citenamefont {Tran}\ \emph {et~al.}(2022)\citenamefont {Tran}, \citenamefont {Zhang}, \citenamefont {Morin}, \citenamefont {Chang}, \citenamefont {Barik}, \citenamefont {Yuan}, \citenamefont {Lee}, \citenamefont {Kim}, \citenamefont {Malik}, \citenamefont {Zhang} \emph {et~al.}}]{tran2022extending}%
  \BibitemOpen
  \bibfield  {author} {\bibinfo {author} {\bibfnamefont {M.~A.}\ \bibnamefont {Tran}}, \bibinfo {author} {\bibfnamefont {C.}~\bibnamefont {Zhang}}, \bibinfo {author} {\bibfnamefont {T.~J.}\ \bibnamefont {Morin}}, \bibinfo {author} {\bibfnamefont {L.}~\bibnamefont {Chang}}, \bibinfo {author} {\bibfnamefont {S.}~\bibnamefont {Barik}}, \bibinfo {author} {\bibfnamefont {Z.}~\bibnamefont {Yuan}}, \bibinfo {author} {\bibfnamefont {W.}~\bibnamefont {Lee}}, \bibinfo {author} {\bibfnamefont {G.}~\bibnamefont {Kim}}, \bibinfo {author} {\bibfnamefont {A.}~\bibnamefont {Malik}}, \bibinfo {author} {\bibfnamefont {Z.}~\bibnamefont {Zhang}},  \emph {et~al.},\ }\bibfield  {title} {\enquote {\bibinfo {title} {Extending the spectrum of fully integrated photonics to submicrometre wavelengths},}\ }\href@noop {} {\bibfield  {journal} {\bibinfo  {journal} {Nature}\ }\textbf {\bibinfo {volume} {610}},\ \bibinfo {pages} {54--60} (\bibinfo {year} {2022})}\BibitemShut {NoStop}%
\bibitem [{\citenamefont {Wang}\ \emph {et~al.}(2024{\natexlab{c}})\citenamefont {Wang}, \citenamefont {Zhong}, \citenamefont {Bruns}, \citenamefont {Liang},\ and\ \citenamefont {Dai}}]{wang2024vivo}%
  \BibitemOpen
  \bibfield  {author} {\bibinfo {author} {\bibfnamefont {F.}~\bibnamefont {Wang}}, \bibinfo {author} {\bibfnamefont {Y.}~\bibnamefont {Zhong}}, \bibinfo {author} {\bibfnamefont {O.}~\bibnamefont {Bruns}}, \bibinfo {author} {\bibfnamefont {Y.}~\bibnamefont {Liang}}, \ and\ \bibinfo {author} {\bibfnamefont {H.}~\bibnamefont {Dai}},\ }\bibfield  {title} {\enquote {\bibinfo {title} {In vivo nir-ii fluorescence imaging for biology and medicine},}\ }\href@noop {} {\bibfield  {journal} {\bibinfo  {journal} {Nature Photonics}\ }\textbf {\bibinfo {volume} {18}},\ \bibinfo {pages} {535--547} (\bibinfo {year} {2024}{\natexlab{c}})}\BibitemShut {NoStop}%
\bibitem [{\citenamefont {Newman}\ \emph {et~al.}(2019)\citenamefont {Newman}, \citenamefont {Maurice}, \citenamefont {Drake}, \citenamefont {Stone}, \citenamefont {Briles}, \citenamefont {Spencer}, \citenamefont {Fredrick}, \citenamefont {Li}, \citenamefont {Westly}, \citenamefont {Ilic} \emph {et~al.}}]{newman2019architecture}%
  \BibitemOpen
  \bibfield  {author} {\bibinfo {author} {\bibfnamefont {Z.~L.}\ \bibnamefont {Newman}}, \bibinfo {author} {\bibfnamefont {V.}~\bibnamefont {Maurice}}, \bibinfo {author} {\bibfnamefont {T.}~\bibnamefont {Drake}}, \bibinfo {author} {\bibfnamefont {J.~R.}\ \bibnamefont {Stone}}, \bibinfo {author} {\bibfnamefont {T.~C.}\ \bibnamefont {Briles}}, \bibinfo {author} {\bibfnamefont {D.~T.}\ \bibnamefont {Spencer}}, \bibinfo {author} {\bibfnamefont {C.}~\bibnamefont {Fredrick}}, \bibinfo {author} {\bibfnamefont {Q.}~\bibnamefont {Li}}, \bibinfo {author} {\bibfnamefont {D.}~\bibnamefont {Westly}}, \bibinfo {author} {\bibfnamefont {B.~R.}\ \bibnamefont {Ilic}},  \emph {et~al.},\ }\bibfield  {title} {\enquote {\bibinfo {title} {Architecture for the photonic integration of an optical atomic clock},}\ }\href@noop {} {\bibfield  {journal} {\bibinfo  {journal} {Optica}\ }\textbf {\bibinfo {volume} {6}},\ \bibinfo {pages} {680--685} (\bibinfo {year} {2019})}\BibitemShut {NoStop}%
\bibitem [{\citenamefont {Cheng}\ \emph {et~al.}(2018)\citenamefont {Cheng}, \citenamefont {Bahadori}, \citenamefont {Glick}, \citenamefont {Rumley},\ and\ \citenamefont {Bergman}}]{cheng2018recent}%
  \BibitemOpen
  \bibfield  {author} {\bibinfo {author} {\bibfnamefont {Q.}~\bibnamefont {Cheng}}, \bibinfo {author} {\bibfnamefont {M.}~\bibnamefont {Bahadori}}, \bibinfo {author} {\bibfnamefont {M.}~\bibnamefont {Glick}}, \bibinfo {author} {\bibfnamefont {S.}~\bibnamefont {Rumley}}, \ and\ \bibinfo {author} {\bibfnamefont {K.}~\bibnamefont {Bergman}},\ }\bibfield  {title} {\enquote {\bibinfo {title} {Recent advances in optical technologies for data centers: a review},}\ }\href@noop {} {\bibfield  {journal} {\bibinfo  {journal} {Optica}\ }\textbf {\bibinfo {volume} {5}},\ \bibinfo {pages} {1354--1370} (\bibinfo {year} {2018})}\BibitemShut {NoStop}%
\bibitem [{\citenamefont {Morin}\ \emph {et~al.}(2021)\citenamefont {Morin}, \citenamefont {Chang}, \citenamefont {Jin}, \citenamefont {Li}, \citenamefont {Guo}, \citenamefont {Park}, \citenamefont {Tran}, \citenamefont {Komljenovic},\ and\ \citenamefont {Bowers}}]{morin2021cmos}%
  \BibitemOpen
  \bibfield  {author} {\bibinfo {author} {\bibfnamefont {T.~J.}\ \bibnamefont {Morin}}, \bibinfo {author} {\bibfnamefont {L.}~\bibnamefont {Chang}}, \bibinfo {author} {\bibfnamefont {W.}~\bibnamefont {Jin}}, \bibinfo {author} {\bibfnamefont {C.}~\bibnamefont {Li}}, \bibinfo {author} {\bibfnamefont {J.}~\bibnamefont {Guo}}, \bibinfo {author} {\bibfnamefont {H.}~\bibnamefont {Park}}, \bibinfo {author} {\bibfnamefont {M.~A.}\ \bibnamefont {Tran}}, \bibinfo {author} {\bibfnamefont {T.}~\bibnamefont {Komljenovic}}, \ and\ \bibinfo {author} {\bibfnamefont {J.~E.}\ \bibnamefont {Bowers}},\ }\bibfield  {title} {\enquote {\bibinfo {title} {Cmos-foundry-based blue and violet photonics},}\ }\href@noop {} {\bibfield  {journal} {\bibinfo  {journal} {Optica}\ }\textbf {\bibinfo {volume} {8}},\ \bibinfo {pages} {755--756} (\bibinfo {year} {2021})}\BibitemShut {NoStop}%
\bibitem [{\citenamefont {Suh}\ \emph {et~al.}(2019)\citenamefont {Suh}, \citenamefont {Yi}, \citenamefont {Lai}, \citenamefont {Leifer}, \citenamefont {Grudinin}, \citenamefont {Vasisht}, \citenamefont {Martin}, \citenamefont {Fitzgerald}, \citenamefont {Doppmann}, \citenamefont {Wang} \emph {et~al.}}]{suh2019searching}%
  \BibitemOpen
  \bibfield  {author} {\bibinfo {author} {\bibfnamefont {M.-G.}\ \bibnamefont {Suh}}, \bibinfo {author} {\bibfnamefont {X.}~\bibnamefont {Yi}}, \bibinfo {author} {\bibfnamefont {Y.-H.}\ \bibnamefont {Lai}}, \bibinfo {author} {\bibfnamefont {S.}~\bibnamefont {Leifer}}, \bibinfo {author} {\bibfnamefont {I.~S.}\ \bibnamefont {Grudinin}}, \bibinfo {author} {\bibfnamefont {G.}~\bibnamefont {Vasisht}}, \bibinfo {author} {\bibfnamefont {E.~C.}\ \bibnamefont {Martin}}, \bibinfo {author} {\bibfnamefont {M.~P.}\ \bibnamefont {Fitzgerald}}, \bibinfo {author} {\bibfnamefont {G.}~\bibnamefont {Doppmann}}, \bibinfo {author} {\bibfnamefont {J.}~\bibnamefont {Wang}},  \emph {et~al.},\ }\bibfield  {title} {\enquote {\bibinfo {title} {Searching for exoplanets using a microresonator astrocomb},}\ }\href@noop {} {\bibfield  {journal} {\bibinfo  {journal} {Nature photonics}\ }\textbf {\bibinfo {volume} {13}},\ \bibinfo {pages} {25--30} (\bibinfo {year} {2019})}\BibitemShut {NoStop}%
\bibitem [{\citenamefont {Bradac}\ \emph {et~al.}(2019)\citenamefont {Bradac}, \citenamefont {Gao}, \citenamefont {Forneris}, \citenamefont {Trusheim},\ and\ \citenamefont {Aharonovich}}]{bradac2019quantum}%
  \BibitemOpen
  \bibfield  {author} {\bibinfo {author} {\bibfnamefont {C.}~\bibnamefont {Bradac}}, \bibinfo {author} {\bibfnamefont {W.}~\bibnamefont {Gao}}, \bibinfo {author} {\bibfnamefont {J.}~\bibnamefont {Forneris}}, \bibinfo {author} {\bibfnamefont {M.~E.}\ \bibnamefont {Trusheim}}, \ and\ \bibinfo {author} {\bibfnamefont {I.}~\bibnamefont {Aharonovich}},\ }\bibfield  {title} {\enquote {\bibinfo {title} {Quantum nanophotonics with group iv defects in diamond},}\ }\href@noop {} {\bibfield  {journal} {\bibinfo  {journal} {Nature communications}\ }\textbf {\bibinfo {volume} {10}},\ \bibinfo {pages} {5625} (\bibinfo {year} {2019})}\BibitemShut {NoStop}%
\bibitem [{\citenamefont {Renaud}\ \emph {et~al.}(2023)\citenamefont {Renaud}, \citenamefont {Assumpcao}, \citenamefont {Joe}, \citenamefont {Shams-Ansari}, \citenamefont {Zhu}, \citenamefont {Hu}, \citenamefont {Sinclair},\ and\ \citenamefont {Loncar}}]{renaud2023sub}%
  \BibitemOpen
  \bibfield  {author} {\bibinfo {author} {\bibfnamefont {D.}~\bibnamefont {Renaud}}, \bibinfo {author} {\bibfnamefont {D.~R.}\ \bibnamefont {Assumpcao}}, \bibinfo {author} {\bibfnamefont {G.}~\bibnamefont {Joe}}, \bibinfo {author} {\bibfnamefont {A.}~\bibnamefont {Shams-Ansari}}, \bibinfo {author} {\bibfnamefont {D.}~\bibnamefont {Zhu}}, \bibinfo {author} {\bibfnamefont {Y.}~\bibnamefont {Hu}}, \bibinfo {author} {\bibfnamefont {N.}~\bibnamefont {Sinclair}}, \ and\ \bibinfo {author} {\bibfnamefont {M.}~\bibnamefont {Loncar}},\ }\bibfield  {title} {\enquote {\bibinfo {title} {Sub-1 volt and high-bandwidth visible to near-infrared electro-optic modulators},}\ }\href@noop {} {\bibfield  {journal} {\bibinfo  {journal} {Nature Communications}\ }\textbf {\bibinfo {volume} {14}},\ \bibinfo {pages} {1496} (\bibinfo {year} {2023})}\BibitemShut {NoStop}%
\bibitem [{\citenamefont {Xue}\ \emph {et~al.}(2023)\citenamefont {Xue}, \citenamefont {Shi}, \citenamefont {Ling}, \citenamefont {Gao}, \citenamefont {Hu}, \citenamefont {Zhang}, \citenamefont {Valentine}, \citenamefont {Wu}, \citenamefont {Staffa}, \citenamefont {Javid},\ and\ \citenamefont {Lin}}]{Xue23}%
  \BibitemOpen
  \bibfield  {author} {\bibinfo {author} {\bibfnamefont {S.}~\bibnamefont {Xue}}, \bibinfo {author} {\bibfnamefont {Z.}~\bibnamefont {Shi}}, \bibinfo {author} {\bibfnamefont {J.}~\bibnamefont {Ling}}, \bibinfo {author} {\bibfnamefont {Z.}~\bibnamefont {Gao}}, \bibinfo {author} {\bibfnamefont {Q.}~\bibnamefont {Hu}}, \bibinfo {author} {\bibfnamefont {K.}~\bibnamefont {Zhang}}, \bibinfo {author} {\bibfnamefont {G.}~\bibnamefont {Valentine}}, \bibinfo {author} {\bibfnamefont {X.}~\bibnamefont {Wu}}, \bibinfo {author} {\bibfnamefont {J.}~\bibnamefont {Staffa}}, \bibinfo {author} {\bibfnamefont {U.~A.}\ \bibnamefont {Javid}}, \ and\ \bibinfo {author} {\bibfnamefont {Q.}~\bibnamefont {Lin}},\ }\bibfield  {title} {\enquote {\bibinfo {title} {Full-spectrum visible electro-optic modulator},}\ }\href {\doibase 10.1364/OPTICA.482667} {\bibfield  {journal} {\bibinfo  {journal} {Optica}\ }\textbf {\bibinfo {volume} {10}},\ \bibinfo {pages} {125--126} (\bibinfo {year} {2023})}\BibitemShut {NoStop}%
\bibitem [{\citenamefont {Sabatti}\ \emph {et~al.}(2024)\citenamefont {Sabatti}, \citenamefont {Kellner}, \citenamefont {Kaufmann}, \citenamefont {Chapman}, \citenamefont {Finco}, \citenamefont {Kuttner}, \citenamefont {Maeder},\ and\ \citenamefont {Grange}}]{Sabatti24}%
  \BibitemOpen
  \bibfield  {author} {\bibinfo {author} {\bibfnamefont {A.}~\bibnamefont {Sabatti}}, \bibinfo {author} {\bibfnamefont {J.}~\bibnamefont {Kellner}}, \bibinfo {author} {\bibfnamefont {F.}~\bibnamefont {Kaufmann}}, \bibinfo {author} {\bibfnamefont {R.~J.}\ \bibnamefont {Chapman}}, \bibinfo {author} {\bibfnamefont {G.}~\bibnamefont {Finco}}, \bibinfo {author} {\bibfnamefont {T.}~\bibnamefont {Kuttner}}, \bibinfo {author} {\bibfnamefont {A.}~\bibnamefont {Maeder}}, \ and\ \bibinfo {author} {\bibfnamefont {R.}~\bibnamefont {Grange}},\ }\bibfield  {title} {\enquote {\bibinfo {title} {Extremely high extinction ratio electro-optic modulator via frequency upconversion to visible wavelengths},}\ }\href {\doibase 10.1364/OL.525733} {\bibfield  {journal} {\bibinfo  {journal} {Opt. Lett.}\ }\textbf {\bibinfo {volume} {49}},\ \bibinfo {pages} {3870--3873} (\bibinfo {year} {2024})}\BibitemShut {NoStop}%
\bibitem [{\citenamefont {Celik}\ \emph {et~al.}(2024)\citenamefont {Celik}, \citenamefont {Ammar}, \citenamefont {Park}, \citenamefont {Stokowski}, \citenamefont {Multani}, \citenamefont {Hwang}, \citenamefont {Gyger}, \citenamefont {Guo}, \citenamefont {Fejer},\ and\ \citenamefont {Safavi-Naeini}}]{AmirSN_456nm}%
  \BibitemOpen
  \bibfield  {author} {\bibinfo {author} {\bibfnamefont {O.~T.}\ \bibnamefont {Celik}}, \bibinfo {author} {\bibfnamefont {N.~Y.}\ \bibnamefont {Ammar}}, \bibinfo {author} {\bibfnamefont {T.}~\bibnamefont {Park}}, \bibinfo {author} {\bibfnamefont {H.~S.}\ \bibnamefont {Stokowski}}, \bibinfo {author} {\bibfnamefont {K.~K.~S.}\ \bibnamefont {Multani}}, \bibinfo {author} {\bibfnamefont {A.~Y.}\ \bibnamefont {Hwang}}, \bibinfo {author} {\bibfnamefont {S.}~\bibnamefont {Gyger}}, \bibinfo {author} {\bibfnamefont {Y.}~\bibnamefont {Guo}}, \bibinfo {author} {\bibfnamefont {M.~M.}\ \bibnamefont {Fejer}}, \ and\ \bibinfo {author} {\bibfnamefont {A.~H.}\ \bibnamefont {Safavi-Naeini}},\ }\bibfield  {title} {\enquote {\bibinfo {title} {Roles of temperature, materials, and domain inversion in high-performance, low-bias-drift thin film lithium niobate blue light modulators},}\ }\href {\doibase 10.1364/OE.538150} {\bibfield  {journal} {\bibinfo  {journal} {Opt. Express}\ }\textbf {\bibinfo {volume} {32}},\ \bibinfo {pages}
  {36160--36170} (\bibinfo {year} {2024})}\BibitemShut {NoStop}%
\bibitem [{\citenamefont {Assumpcao}\ \emph {et~al.}(2024)\citenamefont {Assumpcao}, \citenamefont {Renaud}, \citenamefont {Baradari}, \citenamefont {Zeng}, \citenamefont {De-Eknamkul}, \citenamefont {Xin}, \citenamefont {Shams-Ansari}, \citenamefont {Barton}, \citenamefont {Machielse},\ and\ \citenamefont {Loncar}}]{assumpcao2024thin}%
  \BibitemOpen
  \bibfield  {author} {\bibinfo {author} {\bibfnamefont {D.}~\bibnamefont {Assumpcao}}, \bibinfo {author} {\bibfnamefont {D.}~\bibnamefont {Renaud}}, \bibinfo {author} {\bibfnamefont {A.}~\bibnamefont {Baradari}}, \bibinfo {author} {\bibfnamefont {B.}~\bibnamefont {Zeng}}, \bibinfo {author} {\bibfnamefont {C.}~\bibnamefont {De-Eknamkul}}, \bibinfo {author} {\bibfnamefont {C.}~\bibnamefont {Xin}}, \bibinfo {author} {\bibfnamefont {A.}~\bibnamefont {Shams-Ansari}}, \bibinfo {author} {\bibfnamefont {D.}~\bibnamefont {Barton}}, \bibinfo {author} {\bibfnamefont {B.}~\bibnamefont {Machielse}}, \ and\ \bibinfo {author} {\bibfnamefont {M.}~\bibnamefont {Loncar}},\ }\bibfield  {title} {\enquote {\bibinfo {title} {A thin film lithium niobate near-infrared platform for multiplexing quantum nodes},}\ }\href@noop {} {\bibfield  {journal} {\bibinfo  {journal} {Nature communications}\ }\textbf {\bibinfo {volume} {15}},\ \bibinfo {pages} {1--9} (\bibinfo {year} {2024})}\BibitemShut {NoStop}%
\bibitem [{\citenamefont {Huband}\ \emph {et~al.}(2017)\citenamefont {Huband}, \citenamefont {Keeble}, \citenamefont {Zhang}, \citenamefont {Glazer}, \citenamefont {Bartasyte},\ and\ \citenamefont {Thomas}}]{huband2017relationship}%
  \BibitemOpen
  \bibfield  {author} {\bibinfo {author} {\bibfnamefont {S.}~\bibnamefont {Huband}}, \bibinfo {author} {\bibfnamefont {D.}~\bibnamefont {Keeble}}, \bibinfo {author} {\bibfnamefont {N.}~\bibnamefont {Zhang}}, \bibinfo {author} {\bibfnamefont {A.}~\bibnamefont {Glazer}}, \bibinfo {author} {\bibfnamefont {A.}~\bibnamefont {Bartasyte}}, \ and\ \bibinfo {author} {\bibfnamefont {P.~A.}\ \bibnamefont {Thomas}},\ }\bibfield  {title} {\enquote {\bibinfo {title} {Relationship between the structure and optical properties of lithium tantalate at the zero-birefringence point},}\ }\href@noop {} {\bibfield  {journal} {\bibinfo  {journal} {Journal of Applied Physics}\ }\textbf {\bibinfo {volume} {121}} (\bibinfo {year} {2017})}\BibitemShut {NoStop}%
\bibitem [{\citenamefont {Leidinger}\ \emph {et~al.}(2015)\citenamefont {Leidinger}, \citenamefont {Fieberg}, \citenamefont {Waasem}, \citenamefont {K{\"u}hnemann}, \citenamefont {Buse},\ and\ \citenamefont {Breunig}}]{leidinger2015comparative}%
  \BibitemOpen
  \bibfield  {author} {\bibinfo {author} {\bibfnamefont {M.}~\bibnamefont {Leidinger}}, \bibinfo {author} {\bibfnamefont {S.}~\bibnamefont {Fieberg}}, \bibinfo {author} {\bibfnamefont {N.}~\bibnamefont {Waasem}}, \bibinfo {author} {\bibfnamefont {F.}~\bibnamefont {K{\"u}hnemann}}, \bibinfo {author} {\bibfnamefont {K.}~\bibnamefont {Buse}}, \ and\ \bibinfo {author} {\bibfnamefont {I.}~\bibnamefont {Breunig}},\ }\bibfield  {title} {\enquote {\bibinfo {title} {Comparative study on three highly sensitive absorption measurement techniques characterizing lithium niobate over its entire transparent spectral range},}\ }\href@noop {} {\bibfield  {journal} {\bibinfo  {journal} {Optics express}\ }\textbf {\bibinfo {volume} {23}},\ \bibinfo {pages} {21690--21705} (\bibinfo {year} {2015})}\BibitemShut {NoStop}%
\bibitem [{\citenamefont {Glazer}\ \emph {et~al.}(2010)\citenamefont {Glazer}, \citenamefont {Zhang}, \citenamefont {Bartasyte}, \citenamefont {Keeble}, \citenamefont {Huband},\ and\ \citenamefont {Thomas}}]{glazer2010observation}%
  \BibitemOpen
  \bibfield  {author} {\bibinfo {author} {\bibfnamefont {A.~M.}\ \bibnamefont {Glazer}}, \bibinfo {author} {\bibfnamefont {N.}~\bibnamefont {Zhang}}, \bibinfo {author} {\bibfnamefont {A.}~\bibnamefont {Bartasyte}}, \bibinfo {author} {\bibfnamefont {D.~S.}\ \bibnamefont {Keeble}}, \bibinfo {author} {\bibfnamefont {S.}~\bibnamefont {Huband}}, \ and\ \bibinfo {author} {\bibfnamefont {P.~A.}\ \bibnamefont {Thomas}},\ }\bibfield  {title} {\enquote {\bibinfo {title} {Observation of unusual temperature-dependent stripes in litao3 and litaxnb1- xo3 crystals with near-zero birefringence},}\ }\href@noop {} {\bibfield  {journal} {\bibinfo  {journal} {Journal of Applied Crystallography}\ }\textbf {\bibinfo {volume} {43}},\ \bibinfo {pages} {1305--1313} (\bibinfo {year} {2010})}\BibitemShut {NoStop}%
\bibitem [{\citenamefont {McCracken}, \citenamefont {Charsley},\ and\ \citenamefont {Reid}(2017)}]{mccracken2017decade}%
  \BibitemOpen
  \bibfield  {author} {\bibinfo {author} {\bibfnamefont {R.~A.}\ \bibnamefont {McCracken}}, \bibinfo {author} {\bibfnamefont {J.~M.}\ \bibnamefont {Charsley}}, \ and\ \bibinfo {author} {\bibfnamefont {D.~T.}\ \bibnamefont {Reid}},\ }\bibfield  {title} {\enquote {\bibinfo {title} {A decade of astrocombs: recent advances in frequency combs for astronomy},}\ }\href@noop {} {\bibfield  {journal} {\bibinfo  {journal} {Optics express}\ }\textbf {\bibinfo {volume} {25}},\ \bibinfo {pages} {15058--15078} (\bibinfo {year} {2017})}\BibitemShut {NoStop}%
\bibitem [{\citenamefont {Wang}\ \emph {et~al.}(2022)\citenamefont {Wang}, \citenamefont {Li}, \citenamefont {Yao}, \citenamefont {Li}, \citenamefont {Wu}, \citenamefont {Chiang},\ and\ \citenamefont {Chen}}]{temp_and_voltage_dependence_bias_drift}%
  \BibitemOpen
  \bibfield  {author} {\bibinfo {author} {\bibfnamefont {M.}~\bibnamefont {Wang}}, \bibinfo {author} {\bibfnamefont {J.}~\bibnamefont {Li}}, \bibinfo {author} {\bibfnamefont {H.}~\bibnamefont {Yao}}, \bibinfo {author} {\bibfnamefont {X.}~\bibnamefont {Li}}, \bibinfo {author} {\bibfnamefont {J.}~\bibnamefont {Wu}}, \bibinfo {author} {\bibfnamefont {K.~S.}\ \bibnamefont {Chiang}}, \ and\ \bibinfo {author} {\bibfnamefont {K.}~\bibnamefont {Chen}},\ }\bibfield  {title} {\enquote {\bibinfo {title} {Thin-film lithium-niobate modulator with a combined passive bias and thermo-optic bias},}\ }\href {\doibase 10.1364/OE.474594} {\bibfield  {journal} {\bibinfo  {journal} {Opt. Express}\ }\textbf {\bibinfo {volume} {30}},\ \bibinfo {pages} {39706--39715} (\bibinfo {year} {2022})}\BibitemShut {NoStop}%
\bibitem [{\citenamefont {Zhu}\ \emph {et~al.}(2021{\natexlab{b}})\citenamefont {Zhu}, \citenamefont {Shao}, \citenamefont {Yu}, \citenamefont {Cheng}, \citenamefont {Desiatov}, \citenamefont {Xin}, \citenamefont {Hu}, \citenamefont {Holzgrafe}, \citenamefont {Ghosh}, \citenamefont {Shams-Ansari} \emph {et~al.}}]{zhu2021}%
  \BibitemOpen
  \bibfield  {author} {\bibinfo {author} {\bibfnamefont {D.}~\bibnamefont {Zhu}}, \bibinfo {author} {\bibfnamefont {L.}~\bibnamefont {Shao}}, \bibinfo {author} {\bibfnamefont {M.}~\bibnamefont {Yu}}, \bibinfo {author} {\bibfnamefont {R.}~\bibnamefont {Cheng}}, \bibinfo {author} {\bibfnamefont {B.}~\bibnamefont {Desiatov}}, \bibinfo {author} {\bibfnamefont {C.}~\bibnamefont {Xin}}, \bibinfo {author} {\bibfnamefont {Y.}~\bibnamefont {Hu}}, \bibinfo {author} {\bibfnamefont {J.}~\bibnamefont {Holzgrafe}}, \bibinfo {author} {\bibfnamefont {S.}~\bibnamefont {Ghosh}}, \bibinfo {author} {\bibfnamefont {A.}~\bibnamefont {Shams-Ansari}},  \emph {et~al.},\ }\bibfield  {title} {\enquote {\bibinfo {title} {Integrated photonics on thin-film lithium niobate},}\ }\href@noop {} {\bibfield  {journal} {\bibinfo  {journal} {Advances in Optics and Photonics}\ }\textbf {\bibinfo {volume} {13}},\ \bibinfo {pages} {242--352} (\bibinfo {year} {2021}{\natexlab{b}})}\BibitemShut {NoStop}%
\bibitem [{\citenamefont {Zhu}\ \emph {et~al.}(2024)\citenamefont {Zhu}, \citenamefont {Hu}, \citenamefont {Lu}, \citenamefont {Warner}, \citenamefont {Li}, \citenamefont {Song}, \citenamefont {{a}es}, \citenamefont {Shams-Ansari}, \citenamefont {Cordaro}, \citenamefont {Sinclair},\ and\ \citenamefont {Lon\v{c}ar}}]{AnnaZhu24}%
  \BibitemOpen
  \bibfield  {author} {\bibinfo {author} {\bibfnamefont {X.}~\bibnamefont {Zhu}}, \bibinfo {author} {\bibfnamefont {Y.}~\bibnamefont {Hu}}, \bibinfo {author} {\bibfnamefont {S.}~\bibnamefont {Lu}}, \bibinfo {author} {\bibfnamefont {H.~K.}\ \bibnamefont {Warner}}, \bibinfo {author} {\bibfnamefont {X.}~\bibnamefont {Li}}, \bibinfo {author} {\bibfnamefont {Y.}~\bibnamefont {Song}}, \bibinfo {author} {\bibfnamefont {L.~M.}\ \bibnamefont {{a}es}}, \bibinfo {author} {\bibfnamefont {A.}~\bibnamefont {Shams-Ansari}}, \bibinfo {author} {\bibfnamefont {A.}~\bibnamefont {Cordaro}}, \bibinfo {author} {\bibfnamefont {N.}~\bibnamefont {Sinclair}}, \ and\ \bibinfo {author} {\bibfnamefont {M.}~\bibnamefont {Lon\v{c}ar}},\ }\bibfield  {title} {\enquote {\bibinfo {title} {Twenty-nine million intrinsic q-factor monolithic microresonators on thin-film lithium niobate},}\ }\href {\doibase 10.1364/PRJ.521172} {\bibfield  {journal} {\bibinfo  {journal} {Photon. Res.}\ }\textbf {\bibinfo {volume} {12}},\ \bibinfo {pages} {A63--A68}
  (\bibinfo {year} {2024})}\BibitemShut {NoStop}%
\bibitem [{\citenamefont {Bryan}, \citenamefont {Gerson},\ and\ \citenamefont {Tomaschke}(1984)}]{bryan1984}%
  \BibitemOpen
  \bibfield  {author} {\bibinfo {author} {\bibfnamefont {D.~A.}\ \bibnamefont {Bryan}}, \bibinfo {author} {\bibfnamefont {R.}~\bibnamefont {Gerson}}, \ and\ \bibinfo {author} {\bibfnamefont {H.~E.}\ \bibnamefont {Tomaschke}},\ }\bibfield  {title} {\enquote {\bibinfo {title} {{Increased optical damage resistance in lithium niobate}},}\ }\href {\doibase 10.1063/1.94946} {\bibfield  {journal} {\bibinfo  {journal} {Applied Physics Letters}\ }\textbf {\bibinfo {volume} {44}},\ \bibinfo {pages} {847--849} (\bibinfo {year} {1984})},\ \Eprint {http://arxiv.org/abs/https://pubs.aip.org/aip/apl/article-pdf/44/9/847/18451047/847\_1\_online.pdf} {https://pubs.aip.org/aip/apl/article-pdf/44/9/847/18451047/847\_1\_online.pdf} \BibitemShut {NoStop}%
\bibitem [{\citenamefont {Ashkin}\ \emph {et~al.}(1966)\citenamefont {Ashkin}, \citenamefont {Boyd}, \citenamefont {Dziedzic}, \citenamefont {Smith}, \citenamefont {Ballman}, \citenamefont {Levinstein},\ and\ \citenamefont {Nassau}}]{Bulk_LN_LT_High_Power_Ability}%
  \BibitemOpen
  \bibfield  {author} {\bibinfo {author} {\bibfnamefont {A.}~\bibnamefont {Ashkin}}, \bibinfo {author} {\bibfnamefont {G.~D.}\ \bibnamefont {Boyd}}, \bibinfo {author} {\bibfnamefont {J.~M.}\ \bibnamefont {Dziedzic}}, \bibinfo {author} {\bibfnamefont {R.~G.}\ \bibnamefont {Smith}}, \bibinfo {author} {\bibfnamefont {A.~A.}\ \bibnamefont {Ballman}}, \bibinfo {author} {\bibfnamefont {J.~J.}\ \bibnamefont {Levinstein}}, \ and\ \bibinfo {author} {\bibfnamefont {K.}~\bibnamefont {Nassau}},\ }\bibfield  {title} {\enquote {\bibinfo {title} {{OPTICALLY‐INDUCED REFRACTIVE INDEX INHOMOGENEITIES IN LiNbO3 AND LiTaO3}},}\ }\href {\doibase 10.1063/1.1754607} {\bibfield  {journal} {\bibinfo  {journal} {Applied Physics Letters}\ }\textbf {\bibinfo {volume} {9}},\ \bibinfo {pages} {72--74} (\bibinfo {year} {1966})},\ \Eprint {http://arxiv.org/abs/https://pubs.aip.org/aip/apl/article-pdf/9/1/72/18419070/72\_1\_online.pdf} {https://pubs.aip.org/aip/apl/article-pdf/9/1/72/18419070/72\_1\_online.pdf} \BibitemShut {NoStop}%
\bibitem [{\citenamefont {Wang}\ \emph {et~al.}(2010)\citenamefont {Wang}, \citenamefont {Liu}, \citenamefont {Kong}, \citenamefont {Chen}, \citenamefont {Huang}, \citenamefont {Wu}, \citenamefont {Rupp},\ and\ \citenamefont {Xu}}]{Tin_Bulk_LN}%
  \BibitemOpen
  \bibfield  {author} {\bibinfo {author} {\bibfnamefont {L.}~\bibnamefont {Wang}}, \bibinfo {author} {\bibfnamefont {S.}~\bibnamefont {Liu}}, \bibinfo {author} {\bibfnamefont {Y.}~\bibnamefont {Kong}}, \bibinfo {author} {\bibfnamefont {S.}~\bibnamefont {Chen}}, \bibinfo {author} {\bibfnamefont {Z.}~\bibnamefont {Huang}}, \bibinfo {author} {\bibfnamefont {L.}~\bibnamefont {Wu}}, \bibinfo {author} {\bibfnamefont {R.}~\bibnamefont {Rupp}}, \ and\ \bibinfo {author} {\bibfnamefont {J.}~\bibnamefont {Xu}},\ }\bibfield  {title} {\enquote {\bibinfo {title} {Increased optical-damage resistance in tin-doped lithium niobate},}\ }\href {\doibase 10.1364/OL.35.000883} {\bibfield  {journal} {\bibinfo  {journal} {Opt. Lett.}\ }\textbf {\bibinfo {volume} {35}},\ \bibinfo {pages} {883--885} (\bibinfo {year} {2010})}\BibitemShut {NoStop}%
\bibitem [{\citenamefont {Kong}\ \emph {et~al.}(2007)\citenamefont {Kong}, \citenamefont {Liu}, \citenamefont {Zhao}, \citenamefont {Liu}, \citenamefont {Chen},\ and\ \citenamefont {Xu}}]{Zr_Bulk_LN}%
  \BibitemOpen
  \bibfield  {author} {\bibinfo {author} {\bibfnamefont {Y.}~\bibnamefont {Kong}}, \bibinfo {author} {\bibfnamefont {S.}~\bibnamefont {Liu}}, \bibinfo {author} {\bibfnamefont {Y.}~\bibnamefont {Zhao}}, \bibinfo {author} {\bibfnamefont {H.}~\bibnamefont {Liu}}, \bibinfo {author} {\bibfnamefont {S.}~\bibnamefont {Chen}}, \ and\ \bibinfo {author} {\bibfnamefont {J.}~\bibnamefont {Xu}},\ }\bibfield  {title} {\enquote {\bibinfo {title} {{Highly optical damage resistant crystal: Zirconium-oxide-doped lithium niobate}},}\ }\href {\doibase 10.1063/1.2773742} {\bibfield  {journal} {\bibinfo  {journal} {Applied Physics Letters}\ }\textbf {\bibinfo {volume} {91}},\ \bibinfo {pages} {081908} (\bibinfo {year} {2007})},\ \Eprint {http://arxiv.org/abs/https://pubs.aip.org/aip/apl/article-pdf/doi/10.1063/1.2773742/13977509/081908\_1\_online.pdf} {https://pubs.aip.org/aip/apl/article-pdf/doi/10.1063/1.2773742/13977509/081908\_1\_online.pdf} \BibitemShut {NoStop}%
\bibitem [{\citenamefont {Salvestrini}\ \emph {et~al.}(2011{\natexlab{a}})\citenamefont {Salvestrini}, \citenamefont {Guilbert}, \citenamefont {Fontana}, \citenamefont {Abarkan},\ and\ \citenamefont {Gille}}]{DC_LN_mechanism}%
  \BibitemOpen
  \bibfield  {author} {\bibinfo {author} {\bibfnamefont {J.~P.}\ \bibnamefont {Salvestrini}}, \bibinfo {author} {\bibfnamefont {L.}~\bibnamefont {Guilbert}}, \bibinfo {author} {\bibfnamefont {M.}~\bibnamefont {Fontana}}, \bibinfo {author} {\bibfnamefont {M.}~\bibnamefont {Abarkan}}, \ and\ \bibinfo {author} {\bibfnamefont {S.}~\bibnamefont {Gille}},\ }\bibfield  {title} {\enquote {\bibinfo {title} {Analysis and control of the dc drift in linbo3based mach–zehnder modulators},}\ }\href {\doibase 10.1109/JLT.2011.2136322} {\bibfield  {journal} {\bibinfo  {journal} {Journal of Lightwave Technology}\ }\textbf {\bibinfo {volume} {29}},\ \bibinfo {pages} {1522--1534} (\bibinfo {year} {2011}{\natexlab{a}})}\BibitemShut {NoStop}%
\bibitem [{\citenamefont {Iwasaki}\ \emph {et~al.}(1968)\citenamefont {Iwasaki}, \citenamefont {Yamada}, \citenamefont {Niizeki}, \citenamefont {Toyoda},\ and\ \citenamefont {Kubota}}]{Iwasaki_1968}%
  \BibitemOpen
  \bibfield  {author} {\bibinfo {author} {\bibfnamefont {H.}~\bibnamefont {Iwasaki}}, \bibinfo {author} {\bibfnamefont {T.}~\bibnamefont {Yamada}}, \bibinfo {author} {\bibfnamefont {N.}~\bibnamefont {Niizeki}}, \bibinfo {author} {\bibfnamefont {H.}~\bibnamefont {Toyoda}}, \ and\ \bibinfo {author} {\bibfnamefont {H.}~\bibnamefont {Kubota}},\ }\bibfield  {title} {\enquote {\bibinfo {title} {Refractive indices of litao$_3$ at high temperatures},}\ }\href {\doibase 10.1143/JJAP.7.185} {\bibfield  {journal} {\bibinfo  {journal} {Japanese Journal of Applied Physics}\ }\textbf {\bibinfo {volume} {7}},\ \bibinfo {pages} {185} (\bibinfo {year} {1968})}\BibitemShut {NoStop}%
\bibitem [{\citenamefont {Takigawa}\ \emph {et~al.}(2019)\citenamefont {Takigawa}, \citenamefont {Tomimatsu}, \citenamefont {Higurashi},\ and\ \citenamefont {Asano}}]{Stress_LN}%
  \BibitemOpen
  \bibfield  {author} {\bibinfo {author} {\bibfnamefont {R.}~\bibnamefont {Takigawa}}, \bibinfo {author} {\bibfnamefont {T.}~\bibnamefont {Tomimatsu}}, \bibinfo {author} {\bibfnamefont {E.}~\bibnamefont {Higurashi}}, \ and\ \bibinfo {author} {\bibfnamefont {T.}~\bibnamefont {Asano}},\ }\bibfield  {title} {\enquote {\bibinfo {title} {Residual stress in lithium niobate film layer of lnoi/si hybrid wafer fabricated using low-temperature bonding method},}\ }\href {\doibase 10.3390/mi10020136} {\bibfield  {journal} {\bibinfo  {journal} {Micromachines}\ }\textbf {\bibinfo {volume} {10}} (\bibinfo {year} {2019}),\ 10.3390/mi10020136}\BibitemShut {NoStop}%
\bibitem [{\citenamefont {Baklanova}\ \emph {et~al.}(2018)\citenamefont {Baklanova}, \citenamefont {Solnyshkin}, \citenamefont {Kislova}, \citenamefont {Gudkov}, \citenamefont {Belov}, \citenamefont {Shevyakov}, \citenamefont {Zhukov}, \citenamefont {Kiselev},\ and\ \citenamefont {Malinkovich}}]{pyroelectricity_LN}%
  \BibitemOpen
  \bibfield  {author} {\bibinfo {author} {\bibfnamefont {K.~D.}\ \bibnamefont {Baklanova}}, \bibinfo {author} {\bibfnamefont {A.~V.}\ \bibnamefont {Solnyshkin}}, \bibinfo {author} {\bibfnamefont {I.~L.}\ \bibnamefont {Kislova}}, \bibinfo {author} {\bibfnamefont {S.~I.}\ \bibnamefont {Gudkov}}, \bibinfo {author} {\bibfnamefont {A.~N.}\ \bibnamefont {Belov}}, \bibinfo {author} {\bibfnamefont {V.~I.}\ \bibnamefont {Shevyakov}}, \bibinfo {author} {\bibfnamefont {R.~N.}\ \bibnamefont {Zhukov}}, \bibinfo {author} {\bibfnamefont {D.~A.}\ \bibnamefont {Kiselev}}, \ and\ \bibinfo {author} {\bibfnamefont {M.~D.}\ \bibnamefont {Malinkovich}},\ }\bibfield  {title} {\enquote {\bibinfo {title} {Pyroelectric properties and local piezoelectric response of lithium niobate thin films},}\ }\href {\doibase https://doi.org/10.1002/pssa.201700690} {\bibfield  {journal} {\bibinfo  {journal} {physica status solidi (a)}\ }\textbf {\bibinfo {volume} {215}},\ \bibinfo {pages} {1700690} (\bibinfo {year} {2018})},\ \Eprint
  {http://arxiv.org/abs/https://onlinelibrary.wiley.com/doi/pdf/10.1002/pssa.201700690} {https://onlinelibrary.wiley.com/doi/pdf/10.1002/pssa.201700690} \BibitemShut {NoStop}%
\bibitem [{\citenamefont {Wood}\ \emph {et~al.}(2008)\citenamefont {Wood}, \citenamefont {Daniels}, \citenamefont {Brown},\ and\ \citenamefont {Glazer}}]{LT_birefringence}%
  \BibitemOpen
  \bibfield  {author} {\bibinfo {author} {\bibfnamefont {I.~G.}\ \bibnamefont {Wood}}, \bibinfo {author} {\bibfnamefont {P.}~\bibnamefont {Daniels}}, \bibinfo {author} {\bibfnamefont {R.~H.}\ \bibnamefont {Brown}}, \ and\ \bibinfo {author} {\bibfnamefont {A.~M.}\ \bibnamefont {Glazer}},\ }\bibfield  {title} {\enquote {\bibinfo {title} {Optical birefringence study of the ferroelectric phase transition in lithium niobate tantalate mixed crystals},}\ }\href {\doibase 10.1088/0953-8984/20/23/235237} {\bibfield  {journal} {\bibinfo  {journal} {Journal of Physics: Condensed Matter}\ }\textbf {\bibinfo {volume} {20}},\ \bibinfo {pages} {235237} (\bibinfo {year} {2008})}\BibitemShut {NoStop}%
\bibitem [{\citenamefont {Yan}\ \emph {et~al.}(2019)\citenamefont {Yan}, \citenamefont {Huang}, \citenamefont {Zhou}, \citenamefont {Zhao}, \citenamefont {Li}, \citenamefont {Li}, \citenamefont {Yi}, \citenamefont {Huang}, \citenamefont {Lin}, \citenamefont {Zhang} \emph {et~al.}}]{LT_SAW}%
  \BibitemOpen
  \bibfield  {author} {\bibinfo {author} {\bibfnamefont {Y.}~\bibnamefont {Yan}}, \bibinfo {author} {\bibfnamefont {K.}~\bibnamefont {Huang}}, \bibinfo {author} {\bibfnamefont {H.}~\bibnamefont {Zhou}}, \bibinfo {author} {\bibfnamefont {X.}~\bibnamefont {Zhao}}, \bibinfo {author} {\bibfnamefont {W.}~\bibnamefont {Li}}, \bibinfo {author} {\bibfnamefont {Z.}~\bibnamefont {Li}}, \bibinfo {author} {\bibfnamefont {A.}~\bibnamefont {Yi}}, \bibinfo {author} {\bibfnamefont {H.}~\bibnamefont {Huang}}, \bibinfo {author} {\bibfnamefont {J.}~\bibnamefont {Lin}}, \bibinfo {author} {\bibfnamefont {S.}~\bibnamefont {Zhang}},  \emph {et~al.},\ }\bibfield  {title} {\enquote {\bibinfo {title} {Wafer-scale fabrication of 42° rotated y-cut litao$_3$-on-insulator (ltoi) substrate for a saw resonator},}\ }\href@noop {} {\bibfield  {journal} {\bibinfo  {journal} {ACS Applied Electronic Materials}\ }\textbf {\bibinfo {volume} {1}},\ \bibinfo {pages} {1660--1666} (\bibinfo {year} {2019})}\BibitemShut {NoStop}%
\bibitem [{\citenamefont {Maring}\ \emph {et~al.}(2002)\citenamefont {Maring}, \citenamefont {Tavlykaev}, \citenamefont {Ramaswamy},\ and\ \citenamefont {Kostritskii}}]{Maring02}%
  \BibitemOpen
  \bibfield  {author} {\bibinfo {author} {\bibfnamefont {D.~B.}\ \bibnamefont {Maring}}, \bibinfo {author} {\bibfnamefont {R.~F.}\ \bibnamefont {Tavlykaev}}, \bibinfo {author} {\bibfnamefont {R.~V.}\ \bibnamefont {Ramaswamy}}, \ and\ \bibinfo {author} {\bibfnamefont {S.~M.}\ \bibnamefont {Kostritskii}},\ }\bibfield  {title} {\enquote {\bibinfo {title} {Waveguide instability in litao$_3$},}\ }\href {\doibase 10.1364/JOSAB.19.001575} {\bibfield  {journal} {\bibinfo  {journal} {J. Opt. Soc. Am. B}\ }\textbf {\bibinfo {volume} {19}},\ \bibinfo {pages} {1575--1581} (\bibinfo {year} {2002})}\BibitemShut {NoStop}%
\bibitem [{\citenamefont {Salvestrini}\ \emph {et~al.}(2011{\natexlab{b}})\citenamefont {Salvestrini}, \citenamefont {Guilbert}, \citenamefont {Fontana}, \citenamefont {Abarkan},\ and\ \citenamefont {Gille}}]{salvestrini2011}%
  \BibitemOpen
  \bibfield  {author} {\bibinfo {author} {\bibfnamefont {J.~P.}\ \bibnamefont {Salvestrini}}, \bibinfo {author} {\bibfnamefont {L.}~\bibnamefont {Guilbert}}, \bibinfo {author} {\bibfnamefont {M.}~\bibnamefont {Fontana}}, \bibinfo {author} {\bibfnamefont {M.}~\bibnamefont {Abarkan}}, \ and\ \bibinfo {author} {\bibfnamefont {S.}~\bibnamefont {Gille}},\ }\bibfield  {title} {\enquote {\bibinfo {title} {Analysis and control of the dc drift in linbo $ \_ $\{$3$\}$ $-based mach--zehnder modulators},}\ }\href@noop {} {\bibfield  {journal} {\bibinfo  {journal} {Journal of lightwave technology}\ }\textbf {\bibinfo {volume} {29}},\ \bibinfo {pages} {1522--1534} (\bibinfo {year} {2011}{\natexlab{b}})}\BibitemShut {NoStop}%
\bibitem [{\citenamefont {Kharel}\ \emph {et~al.}(2021)\citenamefont {Kharel}, \citenamefont {Reimer}, \citenamefont {Luke}, \citenamefont {He},\ and\ \citenamefont {Zhang}}]{Kharel:21}%
  \BibitemOpen
  \bibfield  {author} {\bibinfo {author} {\bibfnamefont {P.}~\bibnamefont {Kharel}}, \bibinfo {author} {\bibfnamefont {C.}~\bibnamefont {Reimer}}, \bibinfo {author} {\bibfnamefont {K.}~\bibnamefont {Luke}}, \bibinfo {author} {\bibfnamefont {L.}~\bibnamefont {He}}, \ and\ \bibinfo {author} {\bibfnamefont {M.}~\bibnamefont {Zhang}},\ }\bibfield  {title} {\enquote {\bibinfo {title} {Breaking voltage--bandwidth limits in integrated lithium niobate modulators using micro-structured electrodes},}\ }\href {\doibase 10.1364/OPTICA.416155} {\bibfield  {journal} {\bibinfo  {journal} {Optica}\ }\textbf {\bibinfo {volume} {8}},\ \bibinfo {pages} {357--363} (\bibinfo {year} {2021})}\BibitemShut {NoStop}%
\bibitem [{\citenamefont {Wang}\ \emph {et~al.}(2024{\natexlab{d}})\citenamefont {Wang}, \citenamefont {Xing}, \citenamefont {Ruan}, \citenamefont {Yu}, \citenamefont {Chen}, \citenamefont {Ou},\ and\ \citenamefont {Liu}}]{HWang24}%
  \BibitemOpen
  \bibfield  {author} {\bibinfo {author} {\bibfnamefont {H.}~\bibnamefont {Wang}}, \bibinfo {author} {\bibfnamefont {X.}~\bibnamefont {Xing}}, \bibinfo {author} {\bibfnamefont {Z.}~\bibnamefont {Ruan}}, \bibinfo {author} {\bibfnamefont {J.}~\bibnamefont {Yu}}, \bibinfo {author} {\bibfnamefont {K.}~\bibnamefont {Chen}}, \bibinfo {author} {\bibfnamefont {X.}~\bibnamefont {Ou}}, \ and\ \bibinfo {author} {\bibfnamefont {L.}~\bibnamefont {Liu}},\ }\bibfield  {title} {\enquote {\bibinfo {title} {Optical switch with an ultralow dc drift based on thin-film lithium tantalate},}\ }\href {\doibase 10.1364/OL.531263} {\bibfield  {journal} {\bibinfo  {journal} {Opt. Lett.}\ }\textbf {\bibinfo {volume} {49}},\ \bibinfo {pages} {5019--5022} (\bibinfo {year} {2024}{\natexlab{d}})}\BibitemShut {NoStop}%
\end{thebibliography}%

\end{document}